\documentclass[fleqn,usenatbib]{mnras}

\usepackage{newtxtext,newtxmath}

\usepackage[T1]{fontenc}

\DeclareRobustCommand{\VAN}[3]{#2}
\let\VANthebibliography\thebibliography
\def\thebibliography{\DeclareRobustCommand{\VAN}[3]{##3}\VANthebibliography}

\usepackage{graphicx}	
\usepackage{amsmath}	
\usepackage{eso-pic}
\usepackage{csvsimple}  
\usepackage{dirtytalk}  
\hypersetup{
	colorlinks=true,       
	linkcolor=blue,        
	citecolor=blue,        
	filecolor=magenta,     
	urlcolor=blue         
}
\usepackage{booktabs}   
\usepackage[switch]{lineno}

\usepackage{siunitx}
\DeclareSIUnit \h {\ensuremath{\mathit{h}}}
\DeclareSIUnit \pc {pc}

\title[Constraints on compact objects from DES-SN5YR]{Constraints on compact objects from the Dark Energy Survey five-year supernova sample}

 \AddToShipoutPictureBG*{%
  \AtPageUpperLeft{%
    \hspace{\paperwidth}%
    \raisebox{-\baselineskip}{%
      \makebox[0pt][r]{DES-2024-0853, \;\; FERMILAB-PUB-24-0536-V \;\;}
}}}%

\author[P. Shah et. al.]{
\parbox{\textwidth}{
\Large
P.~Shah,$^{1}$\thanks{E-mail: paul.shah.19@ucl.ac.uk}
T.~M.~Davis,$^{2}$
M.~Vincenzi,$^{3}$
P.~Armstrong,$^{5}$
D.~Brout,$^{7}$
R.~Camilleri,$^{2}$
L.~Galbany,$^{8,9}$
J.~Garc\'ia-Bellido,$^{10}$
M.~S.~S.~Gill,$^{11}$
O.~Lahav,$^{1}$
J.~Lee,$^{12}$
C.~Lidman,$^{13,5}$
A.~M\"oller,$^{14}$
M.~Sako,$^{12}$
B.~O.~S\'anchez,$^{15,16}$
M.~Sullivan,$^{4}$
L.~Whiteway,$^{1}$
P.~Wiseman,$^{4}$
S.~Allam,$^{17}$
M.~Aguena,$^{18}$
S.~Bocquet,$^{19}$
D.~Brooks,$^{1}$
D.~L.~Burke,$^{20,11}$
A.~Carnero~Rosell,$^{21,18}$
L.~N.~da Costa,$^{18}$
M.~E.~S.~Pereira,$^{22}$
S.~Desai,$^{23}$
S.~Dodelson,$^{24,25}$
P.~Doel,$^{1}$
I.~Ferrero,$^{26}$
B.~Flaugher,$^{17}$
J.~Frieman,$^{17,27}$
E.~Gaztanaga,$^{8,3,9}$
D.~Gruen,$^{19}$
R.~A.~Gruendl,$^{28,29}$
G.~Gutierrez,$^{17}$
K.~Herner,$^{17}$
S.~R.~Hinton,$^{2}$
D.~L.~Hollowood,$^{30}$
K.~Honscheid,$^{31,32}$
D.~J.~James,$^{7}$
K.~Kuehn,$^{33,34}$
S.~Lee,$^{35}$
J.~L.~Marshall,$^{36}$
J. Mena-Fern{\'a}ndez,$^{37}$
R.~Miquel,$^{38,39}$
J.~Myles,$^{40}$
A.~Palmese,$^{24}$
A.~Pieres,$^{18,41}$
A.~A.~Plazas~Malag\'on,$^{20,11}$
A.~Roodman,$^{20,11}$
S.~Samuroff,$^{42}$
E.~Sanchez,$^{43}$
I.~Sevilla-Noarbe,$^{43}$
M.~Smith,$^{44}$
E.~Suchyta,$^{45}$
M.~E.~C.~Swanson,$^{28}$
G.~Tarle,$^{46}$
C.~To,$^{31}$
V.~Vikram,$^{6}$
and N.~Weaverdyck$^{47,48}$
\begin{center} (DES Collaboration) \end{center}
}
\vspace{0.4cm}
\\
\textit{Affiliations are listed at the end of the paper}
}

\date{Accepted XXX. Received YYY; in original form ZZZ}

\pubyear{2024}

\begin{document}
\label{firstpage}
\pagerange{\pageref{firstpage}--\pageref{lastpage}}
\maketitle


\begin{abstract}
Gravitational lensing magnification of Type Ia supernovae (SNe Ia) allows information to be obtained about the distribution of matter on small scales. In this paper, we derive limits on the fraction $\alpha$ of the total matter density in compact objects (which comprise stars, stellar remnants, small stellar groupings and primordial black holes) of mass $M > 0.03 M_{\odot}$ over cosmological distances. Using 1,532 SNe Ia from the Dark Energy Survey Year 5 sample (DES-SN5YR) combined with a Bayesian prior for the absolute magnitude $M$, we obtain $\alpha < 0.12$ at the 95\% confidence level after marginalisation over cosmological parameters, lensing due to large-scale structure, and intrinsic non-Gaussianity. Similar results are obtained using priors from the cosmic microwave background, baryon acoustic oscillations and galaxy weak lensing, indicating our results do not depend on the background cosmology. We argue our constraints are likely to be conservative (in the sense of the values we quote being higher than the truth), but discuss scenarios in which they could be weakened by systematics of the order of $\Delta \alpha \sim 0.04$. 
\end{abstract}





\section{Introduction}
\label{sec:intro}
The utility of Type Ia supernovae (SNe Ia) in cosmology arises from the fact that they are empirically standardizable candles, and are bright enough to be observed out to redshift $z \sim 2$. Standardization involves adjusting the apparent magnitudes for the observed SN Ia colour, duration, environment and computed Malmquist bias, and after this the intrinsic dispersion in their luminosities is reduced to just $\sim 10\%$ per SN Ia. This was sufficient for just $\sim 50$ of them to establish the existence of dark energy \citep{Riess1998, Perlmutter1999}.
\par 
Modern datasets, notably Pantheon+ \citep{Scolnic2022} and the Dark Energy Survey 5 Year SN Ia survey \citep{Sanchez2024} (hereafter DES-SN5YR), comprise $\sim 1800$ SNe Ia and can be used to build a detailed Hubble diagram of magnitude versus redshift. This diagram, when paired with an estimation of the statistical and systematic uncertainties, and a theoretical computation of luminosity distances in the cosmological model to be tested, is used to construct a Gaussian likelihood. Thus, constraints on cosmological parameters may be derived in a Bayesian framework, assuming a homogeneous and isotropic universe such that the luminosity distance does not depend on the line of sight. For the latest such constraints derived from DES-SN5YR, see \citet{SNKeyPaper, Camilleri2024}.
\par
Inhomogeneity in foreground matter along the line of sight (LOS) will alter luminosity distances by the action of gravitational lensing. This was originally explained by \citet{Zeldovich1964} and \citet{Dyer1972, Dyer1973, Dyer1981} who showed objects would seem to lie further away if their lines of sight travelled along a region that was under-dense compared to the mean density of the universe. Nevertheless, as gravitational lensing does not create or destroy photons it is common to assert the effects of inhomogeneity \say{average away}, even when the lensing is strong and produces multiple images of the source \citep{Weinberg1976}. Strictly speaking, this is only true at linear order as geometric effects and the non-linear conversion from flux to magnitude generically introduce corrections proportional to the variance of the dispersion of lensing $\sigma_{\rm lens}^2$ across different LOS \citep{Kaiser2016}. Assuming that the lensing is weak and due to smoothly distributed matter, a typical SN Ia at redshift $z=0.5$ may be brightened (if the LOS passes close to an overdensity) or dimmed (if the LOS passes through a void) by $\sim 2.5\%$. Therefore, it is typical to ignore these higher order terms and treat gravitational lensing as an additional statistical noise term to be added to covariance matrix \citep[see for example][]{Vincenzi2024}. The value in common usage is $\sigma_{\rm lens} = 0.055z$ mag as estimated by \citet{Jonsson2010}.
\par
Using lensing as a source of information rather than just a noise term has long been of interest to cosmologists, as lensing is sensitive to total mass rather than the (biased) distribution of luminous matter. Gravitational lensing distorts shapes as well as magnifies, and this distortion (referred to as \textit{shear}) may be statistically measured if the source is extended such as a galaxy. The first detection of the cosmic shear of galaxies was reported in \citet{Bacon2000}. While the shape measurement is noisy and potentially biased by intrinsic alignments and other effects, subsequent large-scale surveys and considerable theoretical machinery to control systematics have been deployed to extract constraints on $S_8$ (a parameter convenient to describe the combination of matter density and matter power spectrum that galaxy lensing is sensitive to) at the $\sim 3\%$ level. However, these values are typically in moderate tension with the observed anistropies of the cosmic microwave background \citep[see][]{CosmoIntertwined22}. 
\par 
SNe Ia are effectively point sources at cosmological distances. Hence their shear cannot be measured, but their magnification may be used instead. However, although SN Ia intrinsic magnitudes have less scatter than other sources, because the lensing due to large-scale structure is typically weak, and strong lensing events are extremely rare \citep[to date, only five have been confirmed, see][]{Rodney2016,Rodney2021, Goobar2023, Pierel2024}, the effect has been difficult to observe. Detections of weak lensing at a significance level of $\sim 1.4\sigma$ were reported by \citet{Jonsson2010,Smith2014, MacAulay2020}, and by \citet{Kronborg2010, Shah2022} at $2.9\sigma$. A first detection at $> 5\sigma$ of the weak lensing of SNe Ia has been made only relatively recently (\citealt{Shah2024}, hereafter S24). 
\par
Nevertheless, many authors have proposed them as an alternative source to probe the distribution of foreground matter, emphasizing the fact that compact sources are uniquely sensitive to the presence of compact lenses (this is discussed further below). An initial exploration of the information that may be extracted by measuring the lensing of SNe Ia was by made by \citet{Refsdal1964}. Using SNe Ia to detect compact dark matter (such as primordial black holes) appears to have been first discussed by \cite{Schneider1987, Linder1988}, and a quantitative prediction of expected constraints was first made by \citet{Metcalf1999}. Primordial black holes (PBH) have long been considered an attractive candidate for dark matter, and remain motivated by observations of the mergers of intermediate-mass black holes \citep{Bird2016, Clesse2017, Sasaki2016}, the continued non-detection of a microscopic candidate for dark matter \citep[for a review, see Chapter 27 of ][]{PDG2022} or deviations of gravity from General Relativity's predictions \citep{Ishak2019}. While numerous constraints on their abundance have been derived in the literature \citep[for a review, see][]{Carr2021}, the differing astrophysical assumptions that these results rely on motivate continued research in this area.
\par
In addition to the search for dark matter constituents, studying SN Ia weak lensing has many benefits for cosmology. Firstly, the SN Ia Hubble diagram is directly affected by lensing assumptions as the computation of Malmquist bias requires an assumed lensing probability density function (pdf). In the case of DES-SN5YR, this was derived from ray-tracing in the MICE N-body simulation \citep{Carretero2015}. The simulations do not include the presence of compact objects, and additionally may not be reliable due to the minimum particle mass (typically $10^9 M_{\odot}$) and softening length (typically $50$ kpc) used. As we will explain further below, were compact objects to be present in significant numbers, the bias corrections - and hence cosmological parameters derived from SNe Ia - would be inaccurate. 
\par 
Secondly, observations of SNe Ia are assumed to be free of bias due to the environment around the line of sight. However, effects such as fibre collisions or crowding might favour over-dense or under-dense regions of the universe. As such the SNe Ia are taken to represent a fair sampling of the matter density of the universe. To remedy these issues, in \citet{Shah2023} a procedure was proposed to estimate the lensing pdf directly from the data, and to \say{de-lens} individual SN Ia as a straightforwardly calculated term in the standardisation process. However, this methodology assumes matter is smoothly clumped and compact objects are not present.
\par
Thirdly, the dispersion of magnitudes caused by lensing, $\sigma_{\rm lens}$, is proportional to an integral over distance and the matter power spectrum (see Eqn. \ref{eq:frieman} below). If the number of compact objects is limited, it is shown below that the lensing of SNe Ia is then most sensitive to matter power at scales between $1 < k <100 h$ Mpc$^{-1}$. This has relevance for the so-called \say{$S_8$-tension} mentioned above, as it has been proposed that the tension may be resolved if matter-power is more suppressed than predicted on these scales \citep{Amon2022}.
\par 
The first SN Ia-derived constraint on the fraction of matter $\alpha = \Omega_{\rm CO} / \Omega_{\rm m}$ comprised of compact objects was $\alpha<0.88$ (all constraints quoted in this paper are 95\% confidence) by \citet{Metcalf2007}. Further progress was made by \citet{Zumalacarregui2018} (hereafter ZS18), who modelled the full-shape of the lensing pdf as a function of $\alpha$ and redshift. To obtain a pdf that was both accurate over cosmological distances and precise enough to avoid resolution issues in N-body simulations, the authors used a hybrid method \citep[]{Kainulainen2009,Kainulainen2011,Kainulainen2011a, Marra2013, Quartin2014} combining simulation with a semi-analytic integration over dark matter halo profiles to capture the effect of smoothly distributed matter. This was convolved with a pdf fitted to ray-tracing calculations of compact objects \citep{Rauch1991}. The authors derived  $\alpha<0.35$ (however, see \citealt{GarciaBellido2017} for caveats to this result). Recently, $\alpha < 0.25$ was derived by \citet{Dhawan2023} using a different methodology relying on peak statistics.
\par
In this paper, we extend the method of ZS18 to address some limitations, and apply it to the Dark Energy Survey 5 Year Type-Ia Supernova data set (DES-SN5YR). We fully marginalise over cosmological parameters (including weak lensing due to large-scale structure), intrinsic non-Gaussianity, and allow for covariance between SNe Ia. We demonstrate that our constraints do not rely solely on the (absence of) high-magnification events, or the use of Bayesian priors from other cosmological probes. We test our methodology on simulated SN Ia datasets. The main result of this paper is to derive a new constraint on compact objects. 
\par
Our paper is organised as follows. In Section \ref{sec:theory} we describe the construction of our model. In Section \ref{sec:data}, we briefly describe the data and in Section \ref{sec:like} we describe our modification of the SN Ia likelihood to incorporate lensing. In Section \ref{sec:results}, we present the results of our analysis which will be a limit on the fraction of matter comprised of compact objects. In Section \ref{sec:summary}, we discuss the implications of our result, and the direction of analysis which will be explored in future papers. We set $c=1$ everywhere.

\section{Theory}
\label{sec:theory}
\subsection{Weak lensing magnification by large-scale structure}
\label{sec:lss}
Strong lensing of SNe Ia by large-scale structure features multiple sources separated on the sky by distances typically of the order of the size of the foreground lens, together with large magnification factors and time delays between images \citep{Rodney2016,Rodney2021, Goobar2023, Pierel2024}. \citet{Morgan2023} used a machine learning algorithm to identify three strongly lensed supernova candidates in the DES survey, however none of these have been positively identified as SN Ia and are not included in the DES-SN5YR dataset. We therefore assume that large-scale structure has generated only weak lensing.
\par
Considering an individual and isolated source, its magnification $A \equiv F/F_0$ is defined as the ratio of the observed flux $F$ to a reference flux $F_0 = L / 4\pi D_{\rm L}^2$ where $D_{\rm L}$ is a luminosity distance and $L$ is the source's intrinsic luminosity. There are two choices of distance for defining $F_0$. The first is to use the \say{filled beam} luminosity distance of a homogeneous and isotropic universe (i.e. Friedmann-Robertson-Walker, or FRW) of the same average matter density (that is, with all inhomogeneity smoothed out), in which case 
\begin{equation}
\label{eq:dlfilled}
D_{\rm L, F} = (1+z) \int_{0}^{z} \frac{dz'}{H(z')} \;\;.
\end{equation}
The second is to define the reference flux as that seen through a narrow bicone surrounding the LOS in which all matter has been removed. As matter focuses light, this \say{empty beam} represents the furthest luminosity distance an object can have in an inhomogeneous universe that retains the same background expansion rate $H(z)$. It is
\begin{equation}
\label{eq:dlempty}
D_{\rm L, E} = (1+z)^2 \int_{0}^{z} \frac{dz'}{(1+z')^2 H(z')} \;\;,
\end{equation}
as first stated by \citet{Dyer1972}\footnote{Although this equation is commonly referred to as the Dyer-Roeder distance, it extends the work of many previous authors. See the discussion in Appendix B of \citet{Kantowski1998} for the historical background.} 
\par
For ease of computation, we calculate using the empty beam reference and convert to the filled beam and change in apparent magnitude at the end. We define the incremental magnification $\mu$ (where subscripts are omitted, we mean empty beam values) and $\mu_{\rm F}$ relative to the filled beam,
\begin{equation}
\label{eq:mudef}
\mu \equiv A-1 \equiv  \mu_{\rm F} + \bar{\mu} \;\;,
\end{equation}
where the average $\bar{\mu}$ is over many sources. The definitions above imply that this average converges in the limit of many sources to 
\begin{equation}
\label{eq:mubar}
\bar{\mu} =  (D_{\rm L, E}(z) / D_{\rm L, F}(z))^2 - 1 \;\;.
\end{equation}
Converting $\mu$ into magnitudes $\Delta m$, the magnification of a given source is then
\begin{equation}
\label{eq:deltam}
\Delta m = -2.5\log_{10}{(1+\mu-\bar{\mu})} \;\;.
\end{equation}
Note that $\mu>0$ but $\Delta m$ can take either sign. 
\par
\begin{figure}
    \centering
    \includegraphics[width=\columnwidth]{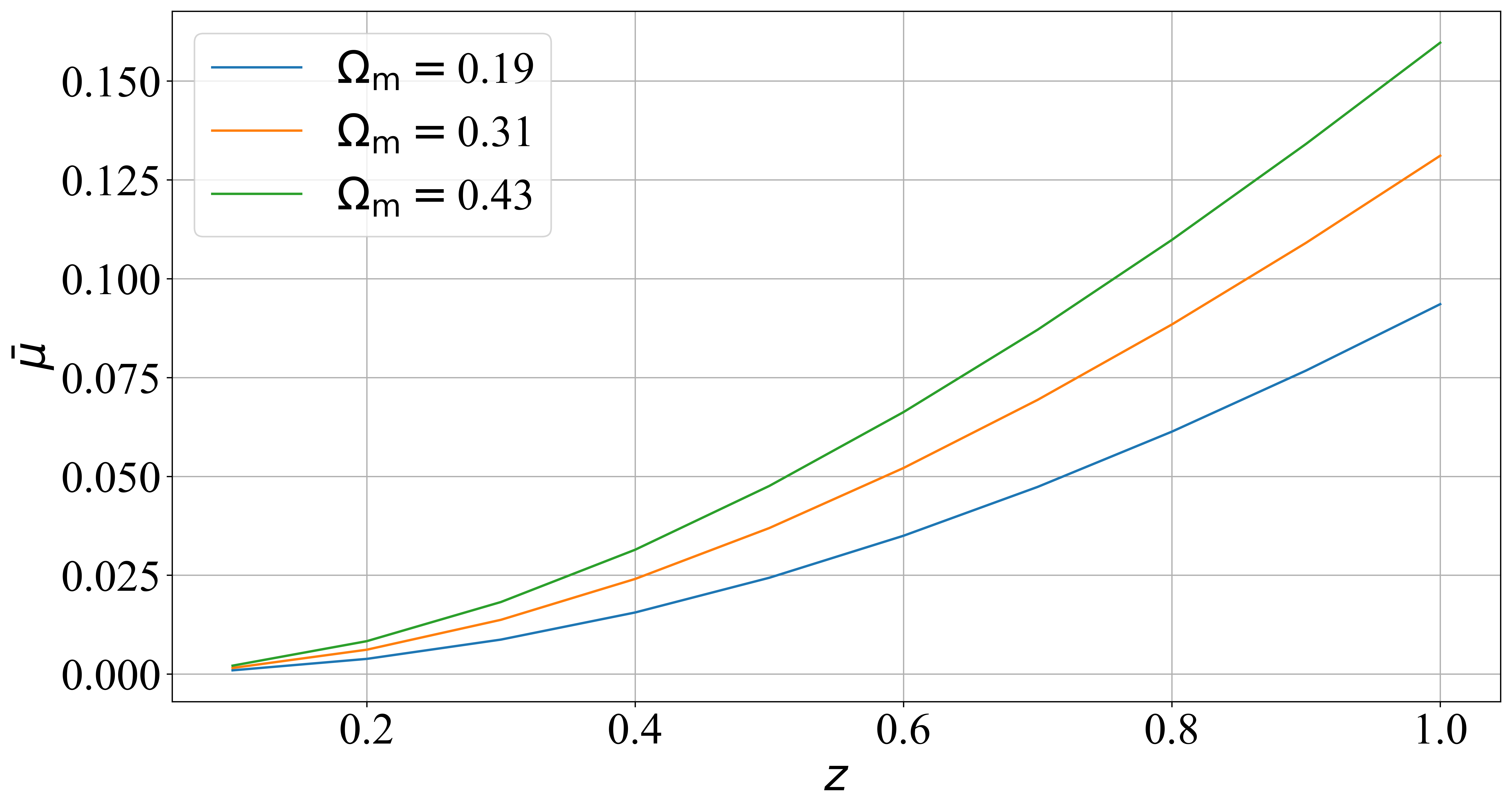}
    \caption{We graph the average empty-beam magnification $\bar{\mu}$ as a function of redshift and $\Omega_{\rm m}$. This quantity represents the average (over many sources) focussing of light by matter compared to a void surrounding the LOS, as specified by $\Omega_{\rm m}$). The units are incremental flux magnification $\mu = F/F_0 -1$ and can be coverted to magnitudes as $\delta m \simeq 1.08 \mu$. Equivalently, it represents the dimming of a source were its line of sight to pass wholly through a void of zero matter density on its journey to the observer. This is then the largest demagnification a source can have in an inhomgeneous universe where the background expansion is unaffected by inhomogeneity.}
    \label{fig:mubar}
\end{figure}

 
The quantity $\bar{\mu}$ is plotted in Figure \ref{fig:mubar}. As we will illustrate later, a redshift-dependent shift in the mode of SNe Ia residuals towards this value, as well as the presence of magnified sources, is a signature of lensing.

\par
We assume a homogeneous spatially flat background cosmology with scale factor $a(\tau)$, and define the Newtonian gauge metric perturbation $\Phi$ to this as
\begin{equation}
    ds^2 = a^2(\tau) [(1+2\Phi) d\tau^2 - (1-2\Phi)(d\chi^2 + \chi^2 (d\theta^2 + \sin^2(\theta) d\phi^2))] \;.
\end{equation}
To linear order, $\nabla^2 \Phi = 4\pi G \langle \rho \rangle \delta$ where the density contrast $\delta = (\rho - \langle \rho \rangle) / \langle \rho \rangle$ for the spatially varying matter density $\rho$, and $a(\tau)$ is determined by the spatial average $\langle \rho \rangle$. We assume there is no \say{back reaction} of inhomogeneities on the background expansion. 
\par
We may define a weak lensing potential $\psi$ as the integral of $\Phi$ over the LOS. The deflection angle of a light ray is then proportional to the product of the gradient in the lens plane of $\psi$ multiplied by a lensing efficiency factor constructed from distances between observer, source and lens. The magnification of an image is then given by the gradient of the deflection angle, now the second derivative of the metric potential. We refer the reader to \citet{Bartelmann2016} for a compact derivation of the relevant equations.
\par
Linearisation in weak lensing occurs in several places: the gravitational potential is assumed small compared to $c$, the inverse Laplacian is localised to the line of sight such that shear contributions can be neglected (this point is relaxed for compact objects), distances are defined as undeflected light paths (known as the Born approximation), and the integral over the LOS implies summation of the small incremental contributions from individual clumps of matter. We find
\begin{equation}
\label{eq:fluxmagnification}
\mu  = \frac{1}{(1-\kappa^2)-|\gamma|^2 }-1 = 2\kappa + \mathcal{O}(\kappa^2, \gamma^2)\;\;,
\end{equation}
where $\gamma$ is the shear. The convergence $\kappa$ is an integral over the comoving distance $\chi$ 
\begin{equation}
\label{eq:kappaempty}
\kappa  \equiv 4\pi G \int_{0}^{\chi_{\rm S}} \frac{\chi (\chi_{\rm S} - \chi)}{\chi_{\rm S}} a^2(\chi) \rho(\chi) d\chi \;\;,
\end{equation}
where $\chi_{\rm S}$ is the comoving distance to the source. 
\par
Eqns. \ref{eq:mudef}, \ref{eq:fluxmagnification} and \ref{eq:kappaempty} then imply that to linear order
\begin{equation}
\label{q:kappafill}
\kappa_{\rm F} = \kappa - \langle \kappa \rangle
\end{equation}
with $\langle \kappa \rangle$ defined by replacing $\rho$ in Eqn. \ref{eq:kappaempty} by $\langle \rho \rangle$.
\par
A point relevant for this paper is the process of linearisation, and in particular the last step of Eqn. \ref{eq:fluxmagnification}, underestimates magnification levels as the neglected higher order terms are positive for all spherically symmetric halo profiles with densities that decline with radius faster than $1/r$. An effect that distinguishes magnification by compact objects as opposed to large-scale structure is that the probability of well-magnified sources is enhanced. As linearisation somewhat underestimates the probability that such events could be instead be caused by large-scale structure, we expect the main result of this paper - which is an upper limit on the presence of compact objects - to be conservative (that is, we quote values that are more likely to be higher than the truth than lower).
\par
It is then straightforward to show \citep[]{Frieman1996} that $\sigma_{\mu}^2 \equiv \langle 4\kappa_{\rm F}^2 \rangle$ can be written as an integral over the power spectrum : 
\begin{equation}
\label{eq:frieman}
\sigma_{\mu}^2 = 9\pi \Omega_{\rm m}^{2} H_0^4 
\int_{0}^{\chi_s} \frac{\chi^2 (\chi_s-\chi)^2}{\chi_s^2} (1+z(\chi))^2 d\chi  \int_{0}^{k_{\rm max}} \frac{\Delta^2(k,z)}{k^2} dk \;\;,
\end{equation}
where $\Delta^2(k,z) =k^3 P(k,z)/2\pi^2$ is the dimensionless matter power spectrum and $\Omega_{\rm m}$ the present day matter density parameter. 
\par
\begin{figure}
    \centering
    \includegraphics[width=\columnwidth]{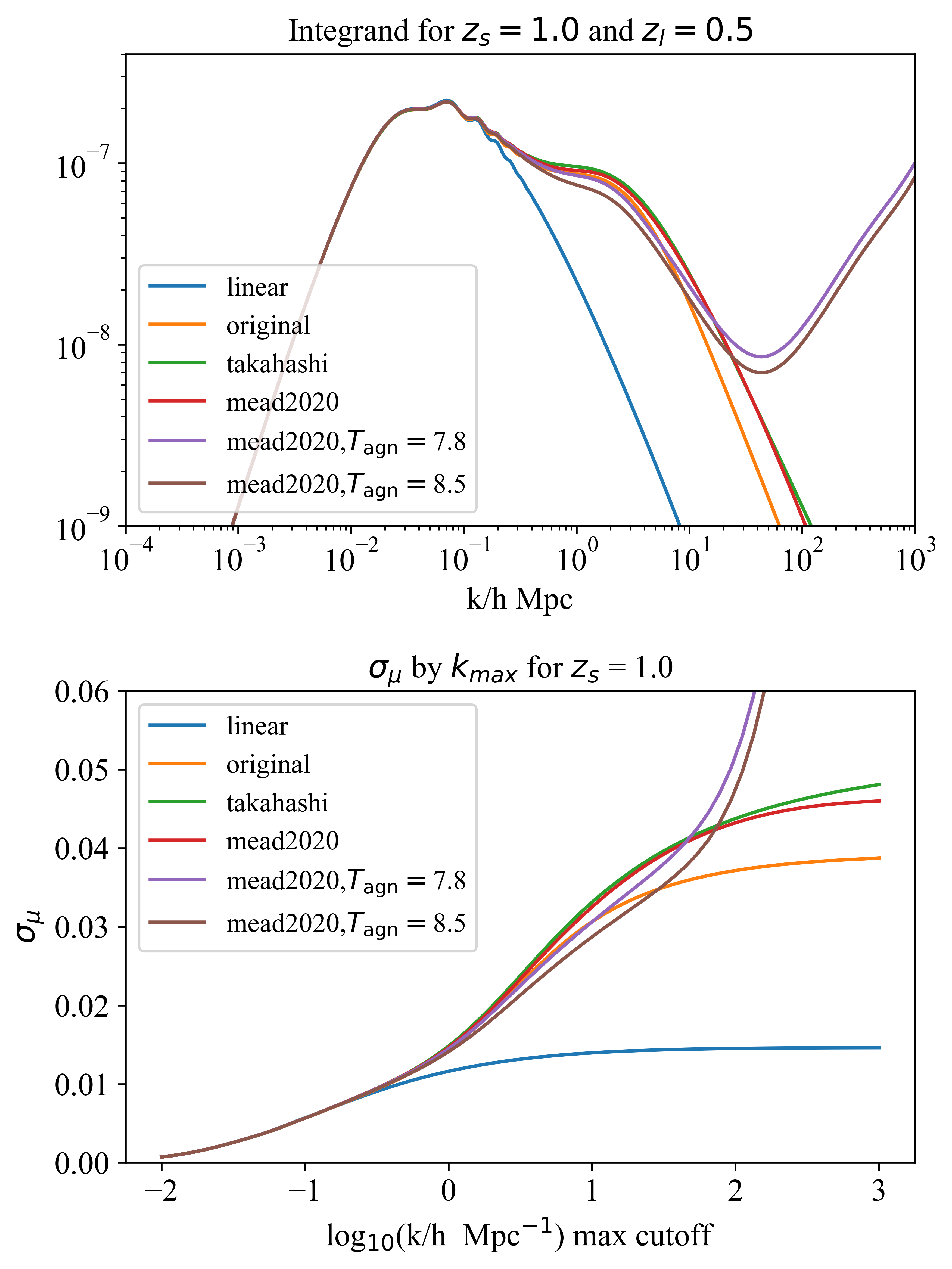} 
    \caption{\textit{Upper panel.} The integrand of Eqn. \ref{eq:frieman} for a lens at $z_l =0.5$ and a source at $z_s = 1.0$, for a range of literature models for the matter power spectrum. The linear power spectrum is shown in blue, the original halo model of \citet{Smith2003} in orange,  \citet{Takahashi2012} in green, the model of \citet{Mead2020} without baryon feedback in red, and with baryon feedback in purple and brown. \textit{Lower panel.} $\sigma_{\mu}$ as a function of the upper cutoff of the integral $k_{\rm max}$ for a source at redshift $z_s = 1$. It is clear the predicted value of $\sigma_\mu$ is sensitive to scales $k/h> 1$ Mpc$^{-1}$. In particular, although models with baryon feedback mildly suppress the dispersion of lensing on intermediate scales, they strongly enhance it from scales $k/h> 10^2$ Mpc$^{-1}$ due to the presence of compact objects. We emphasize that this graph is for illustration only, and the non-linear models and cutoff $k_{\rm max}$ do not have any role in our analysis.}
    \label{fig:sigmu_by_kmax}
\end{figure}

It is informative to consider the behaviour of this integral as a function of the cutoff $k_{\rm max}$ (which could be the source size or beam resolution, and will be large for stellar sources in optical surveys such as those for SNe Ia). To demonstrate this, we consider the power spectrum models provided in \texttt{HMCODE20}\footnote{\url{https://github.com/alexander-mead/HMcode}} \citep{Mead2020}, which combine the linear power spectrum and halo model at non-linear scales, and are calibrated to N-body and hydrodynamical simulations. While the models are not typically used above $k/h >0.2$ Mpc$^{-1}$ (and we do not use them for our results), for the purposes of illustration we use them to graph $\sigma_\mu (k_{\rm max})$ in Figure \ref{fig:sigmu_by_kmax} for a range of parameter options.
Firstly, we see that linear scales contribute little to the total integral. Secondly, the contribution for gravity-only models peaks on scales $1<k/h<100$ Mpc$^{-1}$. Lastly, while baryon feedback (as modelled by \texttt{HMCODE20}) reduces $\sigma_{\mu}^2$ on scales $1<k/h<20$ via active galactic nucleii and supernova feedback, it considerably enhances it on scales $k/h>100$ Mpc$^{-1}$ where baryonic cooling becomes relevant. The integrand is sensitive to the model and assumed parameters, and it is not clear if it is even convergent in the case of baryonic feedback. This also anticipates that problems will occur in deriving lensing pdfs empirically from N-body simulations using particles of size $m_{\rm p} \sim 10^{10} M_{\odot}$ and softening lengths of $ \sim 50$ kpc. In particular, Figure 2 of \citet{Fosalba2015b} shows that the lensing power spectra computed from MICE simulations with differing particle mass (in this case MICE-IR and MICE-GC) start to diverge from each other at multipoles $l > 10^4$, or equivalently $k/h \sim 1$ Mpc$^{-1}$ which is already at the lower end of the $k$ range where the integrand above peaks. 
\par
Figure \ref{fig:sigmu_by_kmax} also implies that using moments of SNe Ia residuals to infer the amplitude of the matter power spectrum, as proposed in \citet{Marra2013, Quartin2014}, may be problematic without a description of small scales.
\par
The situation may seem intractable. However, the following features can be argued from general principles alone : 
\begin{itemize}
\item Lensing shifts the mode of the distribution to demagnification. Voids are larger than lenses, and the majority of lines of sight experience a mild demagnification. Hence there will be larger numbers of dimmed SNe Ia above the Hubble diagram, compared to a small number of brightened SNe Ia below it.
\item The variance of the lensing pdf of large-scale structure increases with distance, the amplitude of the matter power spectrum, and $\Omega_{\rm m}$. These properties are all evident from Eqn. \ref{eq:frieman}.
\item For halo radial density profiles of $\rho(r) \propto r^{-n}$ where $n>2$, the distribution of high $\mu$ values will be dominated by an individual close encounter. The pdf is then driven by max$(\mu_1, \dots, \mu_n)$ rather than the sum \citep{Fleury2020}, and contrary to expectations from the central limit theorem, this skew increases over the redshift range applicable to SN Ia cosmology\footnote{This follows from the Fisher–Tippett–Gnedenko theorem in extreme value theory, which implies convergence to a skewed family of distributions. For an overview of extreme value theory, see \citet{Hansen2020}.}. The distribution of residuals in the presence of lensing can then be expected to have a persistent shape from low to high redshifts, which allows it to be distinguished from observational noise or intrinsic properties.
\end{itemize}
Therefore, it may be argued that the weak lensing pdf due to large-scale structure and haloes has a characteristic and universal shape \citep{Wang2002}, which evolves in a predictable way with changes in cosmological parameters. 
\par
We now turn our attention to scales where baryonic cooling produces compact objects.

\subsection{Lensing magnification by compact objects}
\label{sec:co}
It has long been understood that compact objects (for our paper, we mean this to be an object representable as the Schwarzchild metric) can considerably enhance the values of $\sigma_{\rm lens}$. \citet{Holz2005} obtained $\sigma_{\rm lens} = 0.088z$ by adding compact objects to a model for large-scale structure. In fact, it is straightforward to see that $\sigma_{\rm lens}$ is formally undefined for point sources and point lenses where $p(\mu) \propto \mu^{-3}$ for large $\mu$ \citep{Paczynski1986}. Although cut-offs induced by both the size of the source and lens will ultimately prevent this ultraviolet divergence, this behaviour is very useful observationally : the variance and skew is considerably enhanced compared to large-scale structure.
\par
The lensing pdf due to compact objects has been investigated by a number of authors \citep{Turner1984, Lee1990, Mao1992, Pei1993,  Kofman1997, Lee1997, Fleury2020, Bosca2022}. To make progress, certain assumptions and approximations are usually made (for example, the Born approximation, strongest-lens assumption and so on); it is difficult to analytically derive a form for the pdf that is valid for a three-dimensional distribution of lenses, over a wide range of magnifications. In particular, \citet{Rauch1991} (hereafter R91) gives a fitting formula (Equation 8 of that paper) to numerical results from ray-tracing through an ensemble of randomly-distributed lenses of monochromatic mass $M$. This formula appears to make allowance for the influence of multiple lenses. Alternatively, \citet{Fleury2020} (hereafter F20) use an approximation where only a single, strongest lens interacts with the light ray, which allows them to derive an analytical form of the lensing pdf (Eqn. 23 of that paper). These lenses are also randomly distributed. Finally, \citet{Bosca2022} (hereafter B22) develops this approach to compact objects that follow the same distribution as dark matter (Eqn. 4.20 of that paper). These pdfs all align at low optical depth (defined as the probability of a light ray intersecting the Einstein disc of a lens), and magnifications $\mu<1$, but differ somewhat at high magnifications where the pdf of R91 may understate the probability by a factor of 2 compared to that of B22. We test all three pdfs, but quote our final result using the B22 formula.
\par 
The pdfs have one or two free parameters and are constrained to have the correct normalisation (in the case of F20 and B22 this is guaranteed by construction) and mean. For example, in the case of R91 we have
\begin{equation}
\label{eq:compactobjectpdf}
p_{\rm C}(\mu) = N \left[ \frac{1-\exp{(-\mu/\delta})}{(\mu+1)^2 - 1} \right]^{3/2} \;\;,
\end{equation}
with the parameters $N, \delta$ determined by the constraints 
\begin{equation}
\begin{split}
\int p_{\rm C}(\mu) d\mu & = 1 \\
\int \mu p_{\rm C}(\mu) d\mu & = \bar{\mu}_{\rm C} \;\;.
\end{split}
\end{equation}
If matter is made solely of compact objects, $\bar{\mu}_{\rm C} $ is given by Eqn. \ref{eq:mubar} and for fractional contribution the mean is straightforwardly modified.
\par
A remarkable property is that the pdfs are independent of the mass spectrum of the compact objects. This can be understood qualitatively as follows. The magnification is proportional to the inverse dimensionless impact parameter $R_{\rm E}/b$ where $R_{\rm E}^2 = 4GM D_{\rm LS} D_{\rm L}/D_{\rm S}$ and $b$ is the impact parameter. Hence the magnification scales as $M^{1/2}$ for an individual deflector. However, for a fixed surface mass density, the number of deflectors will scale as $M^{-1/2}$, and the factors of $M$ cancel \citep[see also the discussion in][]{Weinberg1976}. This argument applies only when the source is treated as a point; we discuss finite size sources in Section \ref{sec:caveats}.
\par
A potential source of confusion is that compact objects magnify sources even though the matter density along the line of sight is zero (sometimes referred to as Weyl focussing). This is in apparent contradiction with Eqn. \ref{eq:kappaempty} where weak lensing is proportional to the integral of matter density along the line of sight (sometimes referred to as Ricci focussing). The confusion arises due to the linearisation made to arrive at Eqn. \ref{eq:kappaempty}, from which Ricci focussing arises as a weak lensing limit of the accumulated Weyl focussing of matter inside the beam \citep{Dyer1981}.

\subsection{Modelling SN Ia residuals}
\label{sec:model}
While Eqn. \ref{eq:frieman} above is useful as a guide, it provides us with no information regarding the distribution shape of lensing by large-scale structure. We use instead the code \texttt{TurboGL}\footnote{\url{https://github.com/valerio-marra/turboGL}} \citep{Kainulainen2009, Kainulainen2011a}, with an additional step to incorporate clustering on linear scales.
\par
\texttt{TurboGL} simulates lensing by placing haloes at random distances to the LOS, and calculates the lensing contribution of each halo semi-analytically. The pdf is built from the results of multiple such LOS simulations. This avoids resolution issues discussed above that arise from using pdfs derived from N-body simulations. We use haloes that are spherically symmetric with the standard Navarro-Frenk-White (NFW) profile, and their masses are randomly drawn from the mass function of \citet{Sheth2001} with a lower cut at halo mass $10^7 M_{\odot}$. Spherical symmetry has been shown to be a good approximation when averaged over many LOS \citep{Mandelbaum2005}, and it was shown in S24 that the NFW model is consistent with observations of SN Ia lensing. The halo mass function and profile have been shown to be valid over a wide range halo masses \citep{Wang2020, Zheng2024} encompassing ours. 
\par
The behaviour of the lensing pdf with cosmological parameters arises from distances $\chi(z, \Omega_{\rm m})$, maximum demagnification $\bar{\mu}(z, \Omega_{\rm m})$, linear-scale variance $\sigma_{\mu, \rm Lin}^2 (\Omega_{\rm m}, A_s)$, and the halo mass function $n(M, \Omega_{\rm m}, A_s) dM$ where $A_s$ is the amplitude of the primordial power spectrum. To ensure we are fully able to distinguish lensing by smooth dark matter from compact objects, we generate our pdfs over a range of $(\Omega_{\rm m}, A_s)$ and then marginalise over them.

\subsubsection{Lensing by linear scales}
\texttt{TurboGL} does not incorporate clustering of haloes on linear scales. It has been argued using observed weak lensing maps \citep[for example, see][]{Vikram2015, Jeffrey2021} that the matter density on large scales is reasonably well-approximated as a lognormal distribution \citep{Clerkin2017}. We therefore take the pdf $p_{\rm Lin}(\mu, z ; \Omega_{\rm m}, A_s)$ for linear scales to be described by such with mean $\bar{\mu}$ and variance $\sigma_{\mu, \rm Lin}$ given by Eqn. \ref{eq:frieman} with the linear matter power spectrum. The cutoff $k_{\rm max}$ is not needed as the integral is convergent at large $k$. The pdf is then
\begin{equation}
p_{\rm Lin} = \frac{1}{\mu \sqrt{2\pi B}} \exp{\frac{(\ln{\mu} - C)^2}{2B^2}} \;\;,
\end{equation}
with parameters $B,C$ determined by the constraints 
\begin{equation}
\begin{split}
\int \mu p_{\rm Lin} d\mu  = \bar{\mu} \\
\int (\mu-\bar{\mu})^2 p_{\rm Lin} d\mu  = \sigma_{\mu, \rm Lin}^2 \;\;.
\end{split}
\end{equation}
The second of these two constraints results from the substitution of the linear matter power spectrum $P_{\rm Lin}(k, z ; \Omega_{\rm m}, A_s)$ into Eqn. \ref{eq:frieman}. We have checked our results are unaffected by using a normal distribution with the same mean and variance.

\subsubsection{Lensing by non-linear scales}
\texttt{TurboGL} is run to provide $p_{\rm H}(\mu, z ; \Omega_{\rm m}, A_s)$ for a range of cosmological parameters and redshifts, where the subscript H denotes lensing due to virialised halos. Using the subscript LSS (large-scale structure) to denote the combination of lensing by linear scales and haloes, we form the convolution
\begin{equation}
p_{\rm LSS}(\mu) = \iint  p_{\rm Lin}(\mu_1) p_{\rm H}(\mu_2) \delta(\mu - (\mu_1+ \mu_2)) d\mu_1 d\mu_2 \;\;,
\end{equation}
where $p_{\rm Lin}$ has been adjusted to a mean of zero and $\mu_1$, $\mu_2$ are the contribution of linear scales and haloes respectively to the total lensing magnification $\mu$. The variance of $p_{\rm LSS}$ is then the sum of linear and halo contributions.
\par

\subsubsection{Lensing by compact objects}
The next step is to postulate a fraction $\alpha$ of the matter density is in collapsed objects. We therefore form the total lensing pdf $p_{\rm L}$ by the convolution
\begin{equation}
\begin{split}
p_{\rm L}(\mu) = \iint  p_{\rm LSS}( & (1-\alpha)\mu_1) p_{\rm C}(\alpha \mu_2) \\
& \times \; \delta(\mu - ((1-\alpha)\mu_1+ \alpha \mu_2)) d\mu_1 d\mu_2 \;\;,
\end{split}
\end{equation}
where the variables have been scaled so the distributions $p_{\rm LSS}$ and $p_{\rm C}$ have means of $(1-\alpha)\bar{\mu}$ and $\alpha \bar{\mu}$ respectively.
\par
We pre-compute $p_{\rm L}$ for a sufficiently fine grid of $(z, \mu, \Omega_{\rm m}, A_s, \alpha)$, expressing our cosmology dependence as a scaling with respect to fiducial values $\Omega_{\rm m} = 0.310$ and $A_s = 2.105 \times 10^{-9}$.
\begin{figure}
    \centering
    \includegraphics[width=\columnwidth]{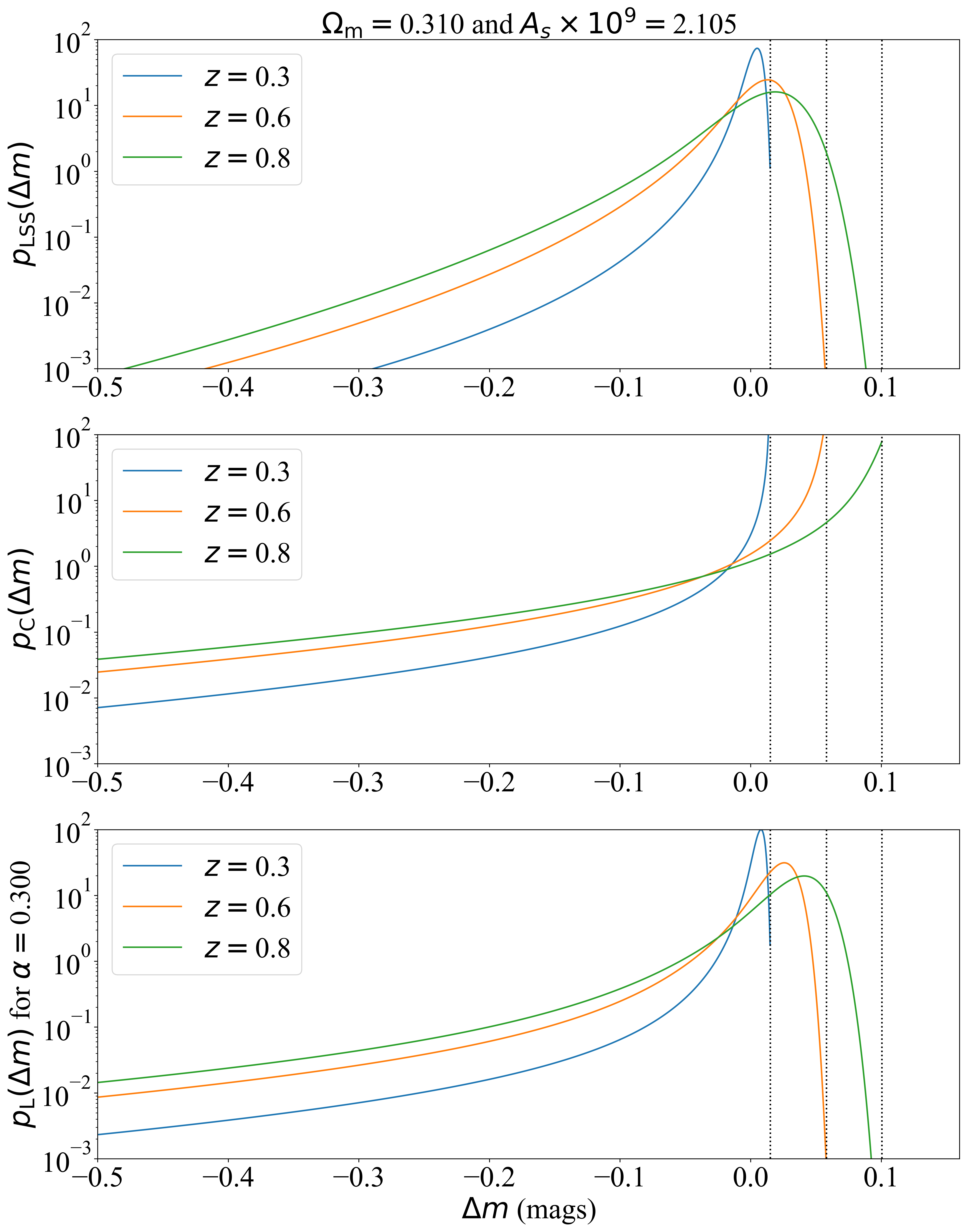}
    \caption{We plot $p_X(\Delta m)$ for a range of redshifts applicable to our data and median values of cosmological parameters. The vertical dashed lines denote the Dyer-Roeder empty-beam distance, which is the maximum demagnification along a LOS devoid of smooth lenses or compact objects. \textit{Upper panel.} The pdf $p_{\rm LSS}$ for weak lensing by large-scale structure only. \textit{Middle panel.} The pdf $p_{\rm C}$ for lensing by compact objects only. \textit{Lower panel.} The combined pdf $p_{\rm L}$ for lensing by 70\% large-scale structure and 30\% compact objects.}
    \label{fig:plss}
\end{figure}
\par
\begin{figure}
    \centering
    \includegraphics[width=\columnwidth]{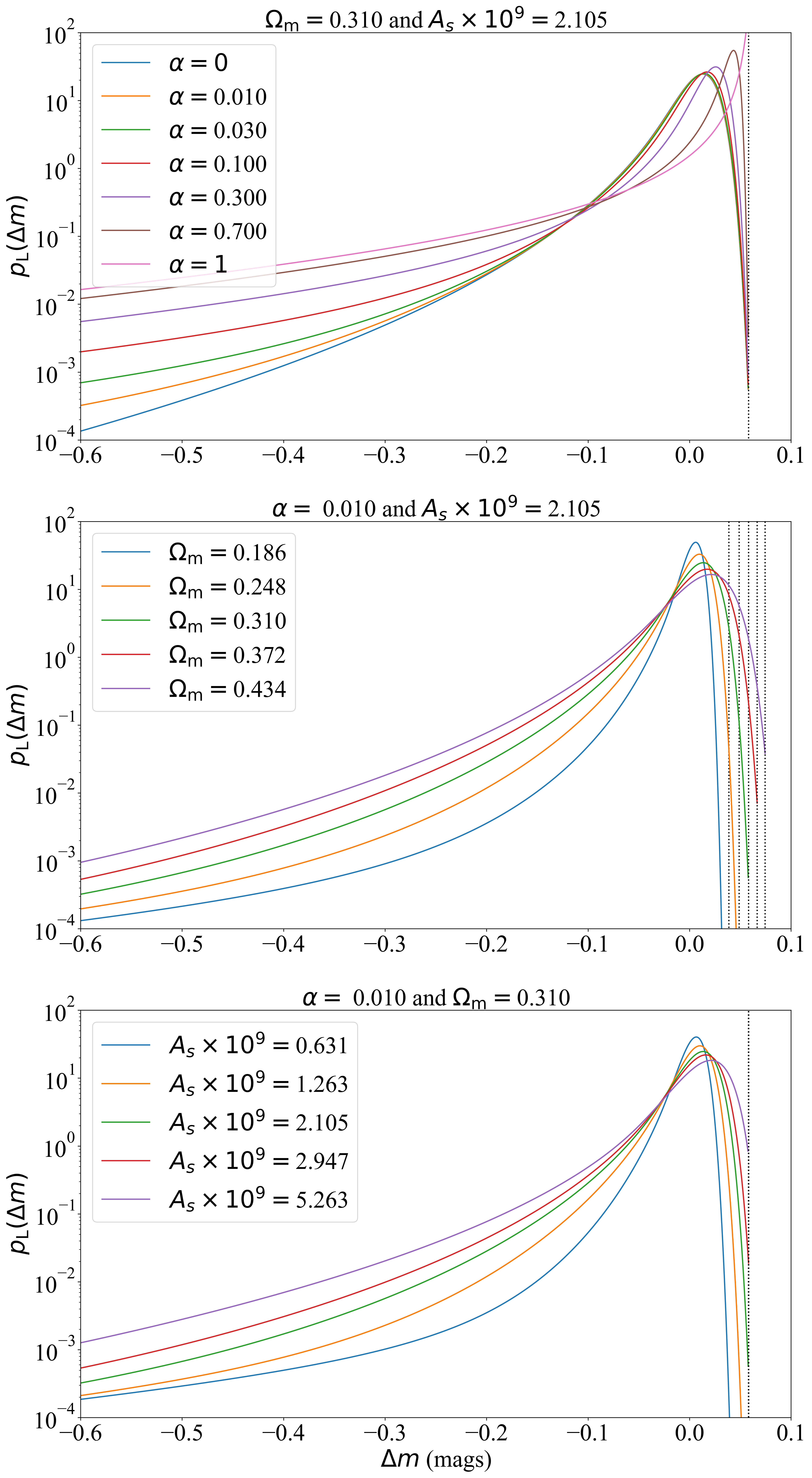}
    \caption{We plot the combined lensing pdf $p_{\rm L}(\Delta m)$ for variations in the fraction of compact objects and cosmological parameters at $z=0.6$. \textit{Upper panel.} Increasing the fraction of compact objects produces a large enhancement in the probability of well-magnified sources. Additionally, the mode of the distribution approaches the Dyer-Roeder empty-beam distance (vertical black dashed line). \textit{Middle panel.} Increasing $\Omega_{\rm m}$ broadens the distribution, in a similar way to the primordial power spectrum amplitude in the lower panel. However, the probability of well-magnified sources is only mildly enhanced compared with variations in the compact object fraction. \textit{Lower panel.} Increasing the amplitude $A_s$ of the primordial power spectrum broadens the distribution. Differences between the middle and lower panels for the probabilities of demagnified events are due to changes in the Dyer-Roeder distance with $\Omega_{\rm m}$.}
    \label{fig:pl}
\end{figure}
\begin{figure}
    \centering
    \includegraphics[width=\columnwidth]{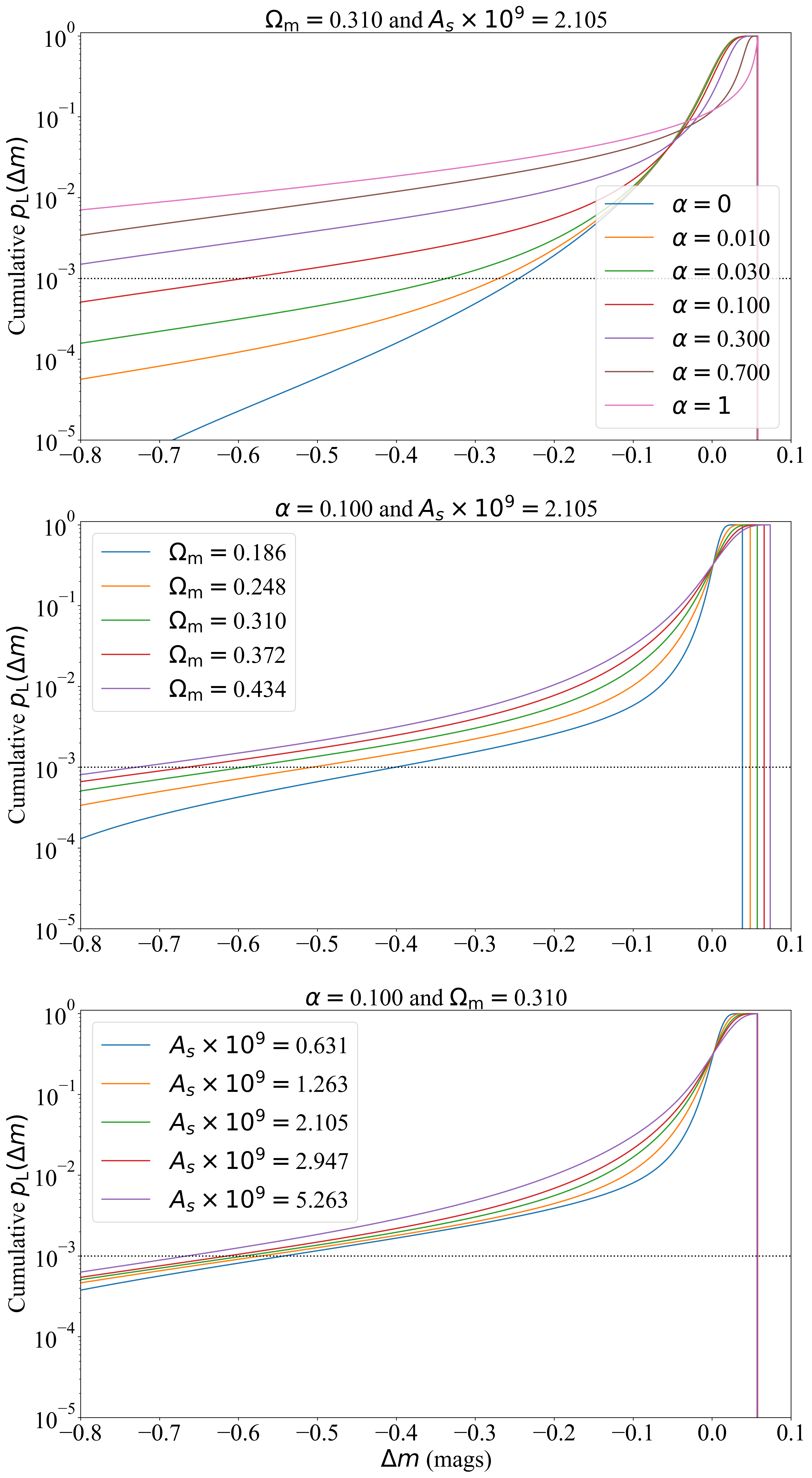}
    \caption{The cumulative probability distribution of the distributions given in Fig. \ref{fig:pl}. As the size of our data is $\sim 1500$ SNe Ia, the intercept of each line with a cumulative probability of $10^{-3}$ gives a guide of the $\Delta m$ to which we may expect approximately one SN Ia to be magnified (before allowing for intrinsic variations). It can also be estimated where larger datasets may produce constraints. }
    \label{fig:cpl}
\end{figure}

We illustrate our large-scale structure, compact object and combined lensing pdfs for a selection of redshifts in Figure \ref{fig:plss}. Considering now the combined pdf, its variation for a range of cosmological parameters is shown in Figures \ref{fig:pl} and \ref{fig:cpl}. In practice, $\Omega_{\rm m}$ will be constrained by the Hubble diagram itself, breaking the partial degeneracy of Eqn. \ref{eq:frieman}. 

\subsubsection{Intrinsic variation} 
The final stage is to convolve with a model of intrinsic SN Ia variation. There is good evidence the SN Ia population is \textit{not} intrinsically Gaussian \citep{Wiseman2022}. While the detail of how this arises is not important for our purposes, we will want to distinguish intrinsic skew (which we assume does not vary with redshift) from skew originating from lensing (which does). A convenient way to parameterize intrinsic skew is using the sin-arcsin distribution family \citep{Jones2009} - the two parameters $\delta, \epsilon$ of this family capture both skew and kurtosis (fat or thin-shouldered distributions) in a monotonic fashion, and $\delta=1, \epsilon =0$ is the normal distribution. The probability of the intrinsic Hubble diagram residuals of supernova $i$ (in terms of magnitudes) is set to 
\begin{equation}
p_{\rm Int}(\Delta m ) = \frac{1}{2\pi E} \delta \sqrt{1+x^2} \exp(-x^2/2)/\sqrt{1+\Delta m^2} \;\;,
\end{equation}
where 
\begin{equation}
x  = (\sinh{(\delta \mbox{arcsinh}{(\Delta m)} - \epsilon)} - D)/E \;\;,
\end{equation}
and the location and scale parameters $D,E$ are determined by the constraints
\begin{equation}
\begin{split}
\int \Delta m p_{\rm Int}(\Delta m) d\Delta m  & = 0 \\
\int \Delta m^2 p_{\rm Int}(\Delta m) d\Delta m  & = \sigma_{i}^{2} \;\;.
\end{split}
\end{equation}
Here $\sigma_i^2 = C_{ii}$ is the diagonal of the SN Ia covariance matrix (see Eqn. \ref{eq:snlike} below), which is the statistical uncertainty in the SN Ia distance modulus. Changing variables in the pdf $p_{\rm L}(\mu)$ to magnitudes using Eqn. \ref{eq:deltam} and convolving with the intrinsic pdf, we finally arrive at the model distribution for SN Ia residuals 
\begin{equation}
\begin{split}
\label{eq:pres}
p_{\rm R}(\Delta m) = \iint p_{\rm L}(\Delta m_1) & p_{\rm Int}(\Delta m_2) \\
& \times \; \delta(\Delta m  - (\Delta m_1+ \Delta m_2)) d\Delta m_2 d\Delta m_2 \;\;,
\end{split}
\end{equation}
where $\Delta m_1$ and $\Delta m_2$ are the contribution of lensing and intrinsic components of the total residual such that $\Delta m = \Delta m_1 + \Delta m_2$.

\section{Data}
\label{sec:data}
\subsection{SN Ia data}

We use the DES-SN5YR dataset as described in \citet{Sanchez2024, SNKeyPaper}. The SN Ia survey was conducted in ten fields (eight shallow and two deep) of the DES footprint, and the SNe Ia range from $ 0.01 < z < 1.13$.  Supernova candidates are analysed using machine-learning classifiers \citep{Moeller2020, Qu2021, Moeller2022} whose input is the time series of light curve fluxes in \textit{griz} passbands and host-galaxy redshifts, and the output is the probability of being an SN Ia. The diagonal of the covariance is then adjusted for this probability, down-weighting likely contaminants but not discarding them altogether. The SN Ia redshift is set to be the post-hoc measured spectroscopic redshift of the galaxy that is closest in directional light radius to the SN Ia \citep{Sullivan2006, Qu2023}. 
\par 
SN Ia magnitudes are empirically standardized by their observed colours, durations, host environments and modelled selection biases. In particular, a lensing pdf derived from the MICE-GC simulations \citep{Fosalba2015a, Crocce2015, Fosalba2015b} was one of the inputs used to calculate the selection bias. Additionally, $\sigma_{\rm lens} = 0.055z$ was added to the covariance matrix to account for lensing dispersion. That these differ somewhat from our model pdfs (and each other) can be neglected if we omit SNe Ia with $z>1$ where the contribution to the bias correction of lensing might be signficant. 
\par
We note that the pipeline and data selection as described in Sections 2 and 5.1 of \citet{Vincenzi2024} implements explicit cuts based on photometric parameters to limit contamination by non-SN Ia. These are based on stretch, colour and goodness-of-fit of the light curve; all of these parameters will be unaffected by lensing. However, at a last stage in the standard DES-SN5YR analysis a Chauvenet cut of outliers with $ | \chi | >4\sigma$ is applied. The intent is to reduce contamination by non-SNIa \citep{Vincenzi2024}, however as this risks also cutting magnified SN Ia from our sample we remove the cut. 
\par
There are 1,905 SNe Ia and we include the 1,556 that are between  $0.2 < z< 1.0$ in our lensing analysis. The lower cut is because lensing will not materially affect low redshift SNe Ia, and the lower redshift SNe Ia are from older, heterogeneous surveys with uncertain selection functions. The upper cut is to reduce potential uncertainties due larger bias corrections at high redshifts \citep[for example, see Figure 7 of][]{Vincenzi2024}. We additionally cut likely SN Ia contaminants, by excluding data with uncertainty $\sigma_{m} > 1.0 $ mag, although we have checked this does not materially affect our results and produces similar results to a cut on non-SN Ia probability. Our data therefore comprises 1,532 SNe Ia of average redshift $ z \sim 0.47$, with just over a third at redshift $z>0.6$ where lensing is expected to comprise a significant proportion of the variance of the Hubble diagram residual.
\par
\begin{figure}
    \centering
    \includegraphics[width=\columnwidth]{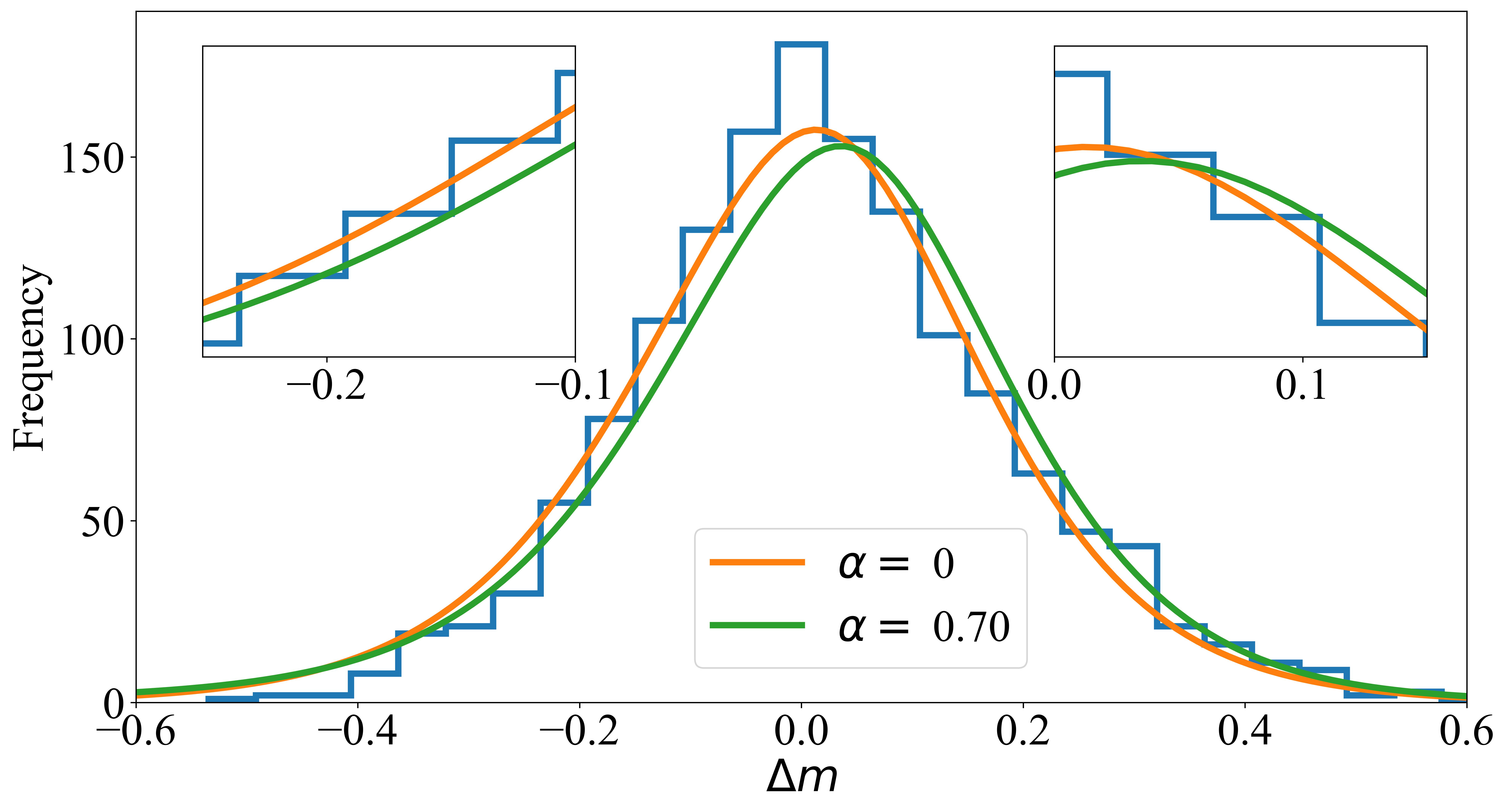}
    \caption{We plot the Hubble diagram residuals $\Delta m = m - m_{\rm theory}$ for SN Ia with $p($SN Ia$) > 0.9$ according to their best-fit cosmology and intrinsic skew and kurtosis. Overlaid on the histogram we illustrate the expected distribution of residuals according to our model pdf given two extreme values of $\alpha = 0$ (orange) and $\alpha = 0.7$ (green). In the two insets, we zoom in on regions of the distribution that will be important in determining our constraints. The differences are not easy to discern due to range of redshifts and unweighted data used for the plot : we caution that $\alpha$ cannot be read off this graph. Nevertheless, it is discernible that $\alpha = 0.7$ is unlikely to be preferred by the data as the peak of the distribution - which moves towards the empty beam value - is away from the peak of the data. Figure \ref{fig:resdist} below shows in greater detail the weighted contribution of individual SN Ia to the lensing likelihood.}
    \label{fig:alphanaddata}
\end{figure}

To aid visualisation of the impact of compact objects, we have fitted a Hubble diagram to our SN Ia data with likely non-SN Ia filtered out and plot a histogram of the residuals $\Delta m = m - m_{\rm theory}$ in Figure \ref{fig:alphanaddata}. We take our background cosmology as Flat $\Lambda$CDM (but see Section \ref{sec:backgroundcosmology} below) and compute $m_{\rm theory} (H_0, \Omega_{\rm m}, z)$ using the standard formulae for distance moduli. Overlaid on the plot is the expected distribution $p_{\rm R}(\Delta m)$ from Eqn. \ref{eq:pres} for large and small values of $\alpha$. The differences are subtle, but nevertheless detectable as is further visualised in Figure \ref{fig:resdist} below.

\subsection{Other cosmological probes}

The amplitude of the primordial power spectrum $A_s$ and $\Omega_{\rm m}$ are constrained by other cosmological probes. In particular, we use:
\begin{itemize}
    \item{Cosmic Microwave Background (CMB) measurements of the temperature and polarisation power spectra (TTTEEE) presented by the \citet{Planck2018}. We use chains derived from the Python implementation of Planck’s 2015 \texttt{Plik\_lite} \citep{prince19}.}
    \item{Weak lensing and galaxy clustering measurements from the DES3$\times$2pt year-3 magnitude-limited (MagLim) lens sample;
    $3\times2$-point refers to the simultaneous fit of three 2-point correlation functions, namely galaxy-galaxy, galaxy-lensing, and lensing-lensing correlations \citep{DES3x2_2022,DES3x2extensions_2023}.}
    \item{Baryon acoustic oscillation (BAO) measurements as presented in the extended Baryon Oscillation Spectroscopic Survey paper \citep[eBOSS;][]{dawson16,alam21}, which adds the BAO results from SDSS-IV \citep{blanton17} to earlier SDSS BAO data. Specifically, we use ``BAO'' to refer to the BAO-only measurements from the Main Galaxy Sample \citep{ross15}, BOSS \citep[SDSS-III][]{alam17}, eBOSS LRG \citep{bautista21}, eBOSS ELG \citep{demattia21}, eBOSS QSO \citep{hou21}, and eBOSS Lya \citep{bourboux20}.
    \item{SH0ES calibration of SN Ia magnitudes by Cepheids. We use $M = -19.253 \pm 0.029$ as specified in Figure 14 of \citet{Riess2022}}}.
\end{itemize}

We use the posterior probabilities and covariance of cosmological parameters $A_s$, $\Omega_{\rm m}$ and $H_0$ from these results as input priors to our SN Ia lensing analysis\footnote{The chains are available at \url{https://github.com/des-science/DES-SN5YR}}. As $A_s$ values are mildly in tension between the CMB and galaxy surveys, we do not combine them but use them individually. We label our first two combinations as SN+BAO+3x2pt and SN+Planck. Our third combination, labelled SN, is to use SNe Ia in combination with the SH0ES calibration of $M$, noting that it is in tension with an inverse distance ladder calibration using the CMB. The purpose of three combinations is to show our results are not sensitive to differences of cosmological parameters between them.

\section{The SN Ia lensing likelihood}
\label{sec:like}

The SN Ia likelihood typically used in cosmology analysis assigns an N-dimensional Gaussian $\mathcal{L}_{G}$ probability to the data $\vec{d}$ given cosmological parameters $\theta$. Hence, up to a normalisation factor 
\begin{equation}
\label{eq:snlike}
\ln \mathcal{L}_{G} (\vec{d} | \theta)  = -\frac{1}{2} d_i C^{-1}_{ij} d_j \equiv -\frac{1}{2} \chi^2  \;\;,
\end{equation}
where $C$ is the covariance matrix (which is the sum of estimated statistical and systematic errors) and the Hubble diagram residuals $d_i$ are 
\begin{equation}
d_i = m_i - \mathcal{M} - \mu_{i, \rm theory}(\theta) \;\;.
\end{equation}
Here $\mathcal{M} = M - 5\log_{10} H_0$ is a degenerate combination of the Hubble constant $H_0$ and the fiducial SN Ia absolute magnitude $M$ (which is marginalised over). The distance modulus $\mu_{i, \rm theory} = 5 \log_{10} (D_{L, F}(z_i, \Omega_{\rm m})/1 \rm{Mpc} ) + 25$ where $D_{L, F}$ is the filled-beam homogeneous luminosity distance as defined in Eqn. \ref{eq:dlfilled}. The observed SNe Ia redshifts are corrected to the CMB rest frame using the heliocentric CMB dipole and a peculiar velocity model. 
\par
We use the pdf for the Hubble diagram residual $p_{\rm R}$ - which combines intrinsic variance and lensing - as specified in Eqn. \ref{eq:pres} for each SN Ia and adjust the above likelihood to 
\begin{equation}
\label{eq:lenslike}
\log \mathcal{L} (d | \theta) = \ln \mathcal{L}_{G}  + \left( \sum_i \log p_{\rm R}(d_i | \theta) - \sum_i \log p_{\rm diag}(d_i | \theta) \right) \;\;.
\end{equation}
The term in brackets is therefore a correction to the standard likelihood incorporating lensing and intrinsic non-Gaussianity. Here 
\begin{equation}
p_{\rm diag}(d_i) = \frac{1}{\sqrt{2\pi} \sigma_i } \exp (-0.5(d_i/\sigma_i)^2) \;\;,
\end{equation}
where $\sigma_i^2 = C_{ii}$ is the diagonal of the covariance matrix which captures the statistical error. 
\par
The covariance matrix $C = C_{\rm stat} + C_{\rm sys}$ is the sum of a diagonal matrix of statistical errors, and an off-diagonal matrix of systematic errors \citep{Vincenzi2024}. The form of our likelihood retains sensitivity to off-diagonal systematics in $C$, which are important to correctly propagate errors. Were $C$ to be fully diagonal, $\log \mathcal{L}$ trivally reduces to the sum of the logs of $p_{\rm R}$ of each $d_i$. Also, in the case of homogeneity when lensing is absent, we may write the lensing pdf $p_{\rm L}$ as a trivial delta function, and $\mathcal{L}$ reduces to the original $\mathcal{L}_{G}$ if intrinsic non-Gaussianity is ignored.
\par
In practice, the covariance matrix $C$ is only weakly non-diagonal, with the typical size of a non-diagonal term being $10^{-3}$ of a diagonal term, so we expect our assignment to be accurate. It is reasonable to neglect covariance between lensing and other sources of uncertainty, and also between lensing and lensing : the SNe Ia are widely enough separated spatially that any covariance induced by overlapping foregrounds is negligible (as tested in S24). 
\par
To speed up computation, we pre-compute the lensing pdfs for a grid of redshifts, cosmological parameters and $\alpha$. We then interpolate the log probabilities by redshift, deviation from the mean and scale parameters. We have checked the interpolation does not affect our results. Only the final convolution with the SN Ia intrinsic pdf $p_{\rm int}$ is calculated in the likelihood. When we do not use the CMB or BAO, our priors (which are also the size of our grid) are $\Omega_{\rm m} \in (0.19, 0.44)$, $10^9 A_s \in (0.63, 5.26)$, $M \in (-18,-21)$ and $H_0 \in (60,80)$. We take $\alpha \in (0.005, 0.7)$ (the lower bound is slightly above zero for numerical stability), $\epsilon \in (-0.2,0.2)$ and $\delta \in (0.6, 1.4)$. Runs are performed using \texttt{Polychord}\footnote{\url{https://github.com/PolyChord/PolyChordLite}} \citep{Polychord}, and plots and analysis with \texttt{Anesthetic}\footnote{\url{https://github.com/handley-lab/anesthetic}} \citep{Anesthetic}.

\section{Results}
\label{sec:results}
For combination of DES-SN5YR and the SH0ES $M$ prior, for the compact objects pdf of B22 we find $\alpha_{\rm SN} < 0.122$ at 95\% confidence after marginalising over $\epsilon, \delta, \Omega_M, \mathcal{M}$ and $A_s$. The median and 68\% confidence levels are $\alpha_{\rm SN}  = 0.033^{+0.045}_{-0.021}$, and the maximum a posteriori value (as determined from kernel density estimation) for $\alpha$ is consistent with zero within error of estimation. Values for $\epsilon, \delta$ are slightly shifted from a normal distribution but not unduly so. 
\par
In combination with Planck priors, we find $\alpha_{\rm SN+CMB} < 0.124$ at 95\% confidence. The median and 68\% confidence levels are $\alpha_{\rm SN+CMB} = 0.032^{+0.047}_{-0.019}$, and the maximum a posteriori is also consistent with zero. 
\par
Finally, when using BAO+3x2pt priors, we find $\alpha_{\rm SN+BAO+3x2pt} < 0.118$ at 95\% confidence, median as $\alpha_{\rm SN+BAO+3x2pt} = 0.033^{+0.045}_{-0.02}$, and the maximum a posteriori is again consistent with zero. The values are summarized in Table 1. We have tested our results using different maximum redshifts and errors in order to verify that the bias correction process (which is redshift dependent) does not influence our results: we find no notable trends. Within statistical fluctuation, these results are all consistent with each other, and we find similar consistency when using the pdfs of R91 and F20 (as expected, since the data prefers a low optical depth). Although the marginalised posterior for $A_s$ for the SN+SH0ES combination peaks at low values, its median and mean is fully consistent with the other datasets. We further discuss the validation of our pipeline and results using simulations in Appendix \ref{sec:validation}.
\par
\begin{figure}
    \centering
    \includegraphics[width=\columnwidth]{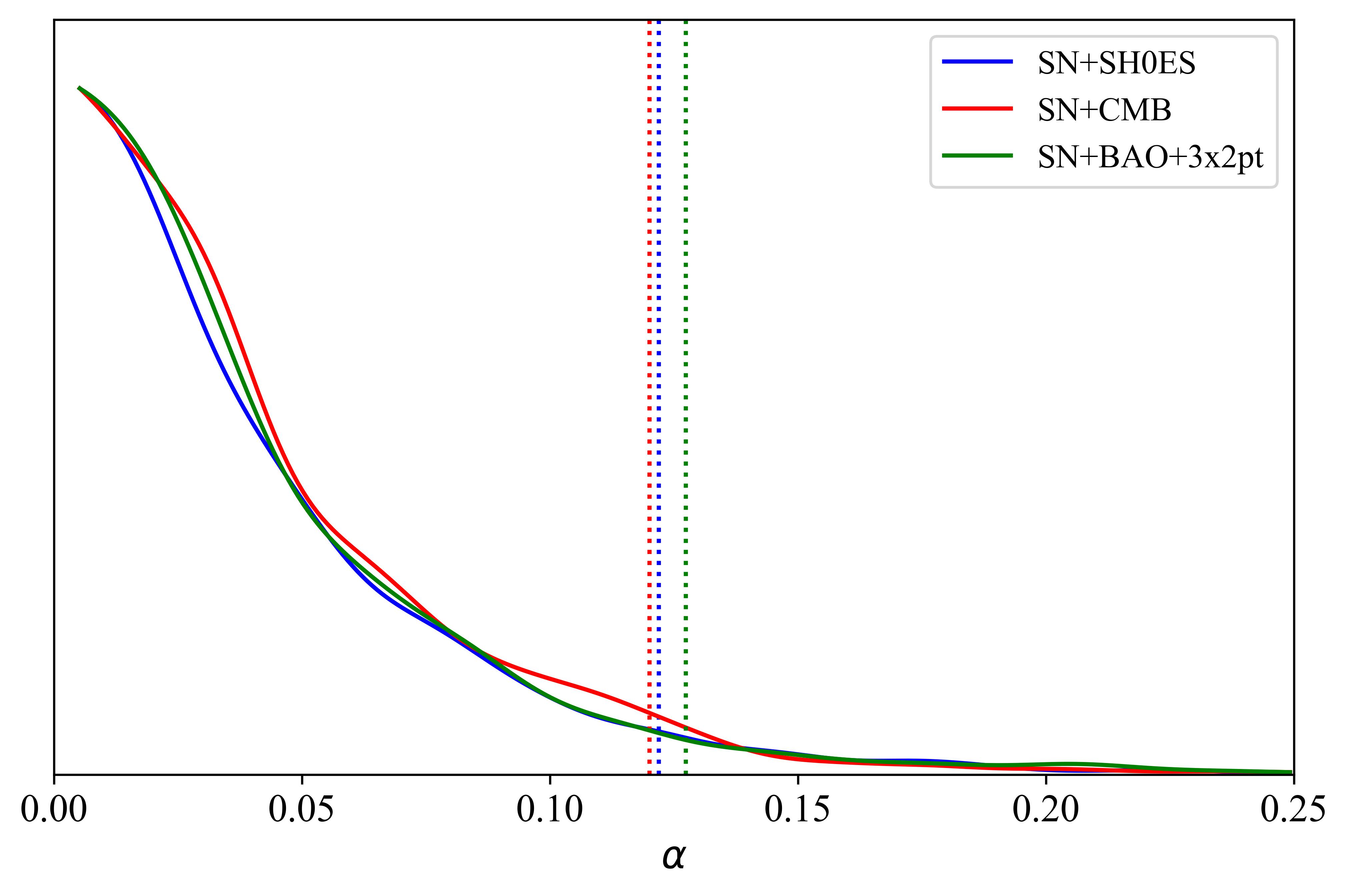}
    \caption{Marginalised model posterior for the fraction of matter $\alpha$ in compact objects, which comprise stars, stellar remnants, stellar groupings below $10^{7} M_{\odot}$ and primordial black holes. The probability density for each data combination has been normalised to a maximum of 1 for plotting purposes, and the 95\% constraints are shown as dotted vertical lines. The median and 68\% confidence interval is shown for the SN+CMB combination. The maximum a posteriori estimates for $\alpha$ are consistent with zero. }
    \label{fig:alpha_post}
\end{figure}
\begin{figure*}
    \centering
    \includegraphics[width=\textwidth]{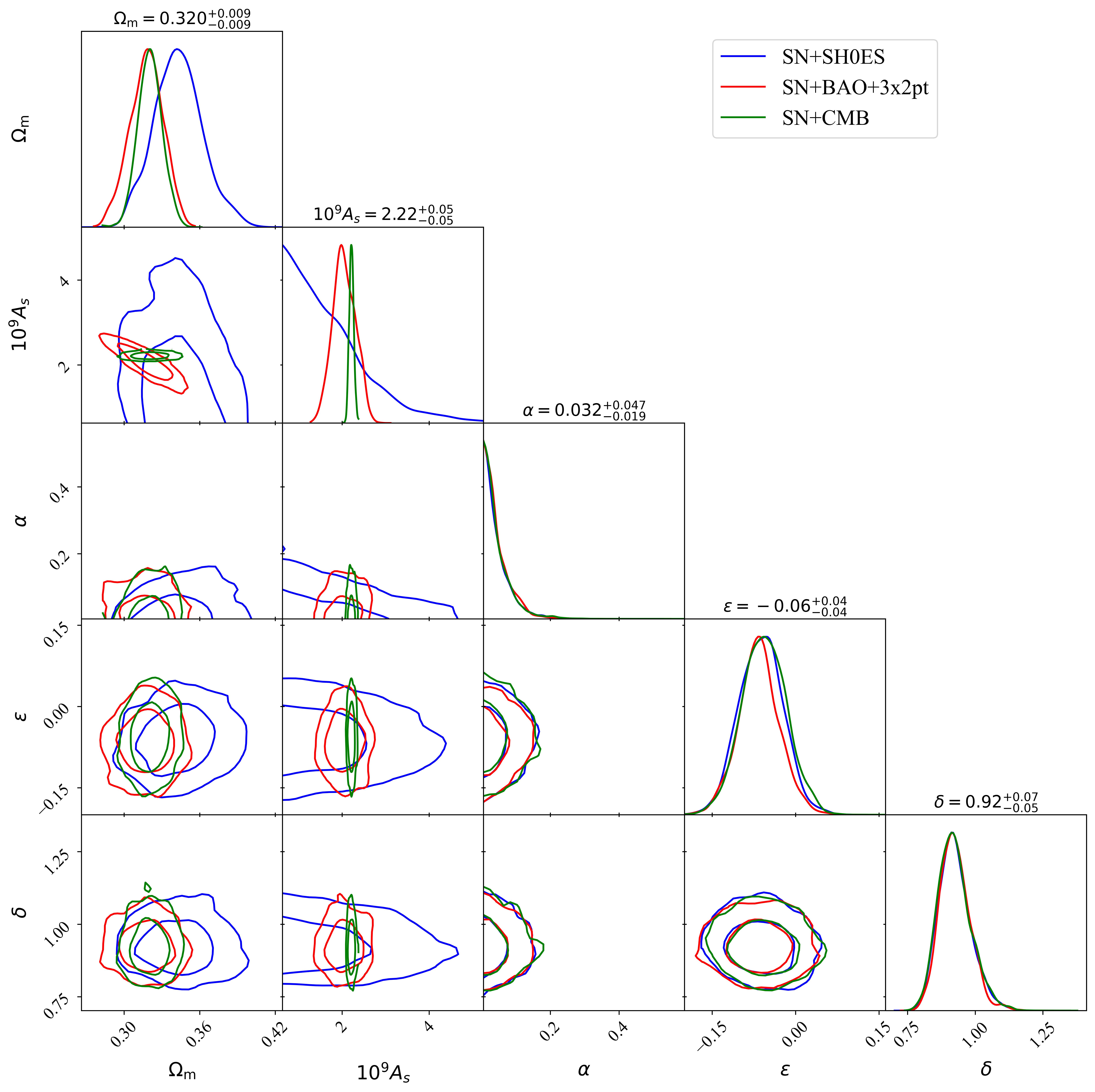}
    \caption{Model posterior parameters for the data combinations used in this paper. For the SN+CMB combination, the medians and 68\% confidence intervals are specified in the text labels above the diagonal. $\alpha$ is the fraction of matter in compact objects, the focus of this paper, and does not show any strong degeneracy with other parameters. The constraints on $A_s$ and $\Omega_{\rm m}$ arise from the combination of the external priors and SN Ia data; however the constraint on $A_s$ from DES-SN5YR alone is weak as the dataset does not have sufficient statistical power to constrain it. $\epsilon$ and $\delta$ represent intrinsic skew and kurtosis of the distribution of SN Ia residuals that is not dependent on redshift, and hence not due to lensing.}
    \label{fig:snbao_triangle}
\end{figure*}

We plot the marginalised posteriors for $\alpha$ of our three data combinations in Figure \ref{fig:alpha_post}, and display the full posteriors in Figure \ref{fig:snbao_triangle}.

\par 
As noted, our posteriors for $\alpha$ peak at values consistent with zero. We compute the Bayes ratio $R = p(M_0 | \vec{d})/p(M_\alpha | \vec{d}) $, which is the relative probabilities of the hypotheses $M_0 : \alpha = \alpha_0$ and $M_\alpha : \alpha \in \pi(\alpha)$ where $\pi$ is our (top-hat) prior for $\alpha$. This may be calculated using the Savage-Dickey ratio
\begin{equation}
    R = \frac{p(\alpha = \alpha_0 | \vec{d}, M_\alpha)}{\pi(\alpha = \alpha_0 | M_\alpha) } \;\;,
\end{equation}
which depends only on the prior and posterior probabilities (as estimated by a kernel density) in $M_\alpha$ localised at $\alpha = \alpha_0$. 
\par
For our uniform prior $\alpha \in (0.005,0.7)$, we find $\ln R = 2.6$ at $\alpha = 0.005$ for all three data combinations. Using the interpretative scale of \citet{Trotta2008}, this indicates weak preference for the absence of a detectable amount of compact objects at odds of 14:1. 

\renewcommand{\arraystretch}{1.5}
\begin{table}
\label{tab:results}
\centering
\begin{tabular}{l r r c} 
 \hline
   & Median($\alpha$) & 95\% upper limit for $\alpha$\\
 \hline 
SN+SH0ES  & $0.033^{+0.045}_{-0.021}$  & 0.122 \\
SN+CMB  & $0.032^{+0.047}_{-0.019}$ & 0.124 \\
SN+BAO+3x2pt  & $0.033^{+0.045}_{-0.02}$ & 0.118 \\
 \hline
\end{tabular}
\caption{Marginalised constraints on $\alpha$, the fraction of the total matter density comprised of compact objects from the data combinations used. For the median 68\% confidence intervals are indicated.}
\end{table}

\subsection{Caveats}
\label{sec:caveats}

In this section, we discuss a range of caveats to compact object constraints that have been put forward in \citet{Carr2024} and (specifically relating to SN Ia lensing) \citet{GarciaBellido2017}. While we leave detailed computations for future work, we estimate in general terms the effect of the points raised by these authors.

\subsubsection{Finite source size}
SN Ia photometry depends on observations around peak brightness. We may estimate the physical size of the photosphere at this peak as $140 $ a.u. from a typical ejecta velocity of $\sim 12,000$ km sec$^{-1}$, and time to peak of $\sim 20$ days. As the mass of the lensing deflector decreases, the angular size of its Einstein radius $\theta_{\rm E}$ approaches the angular size of the SN Ia $\theta_{\rm S}$. Defining the ratio of these sizes as $\eta$, we have
\begin{equation}
\label{eq:eta}
\eta \equiv \frac{\theta_{\rm S}}{\theta_{\rm E}} = 1.52 \left( \frac{M}{M_{\odot}} \right)^{-1/2} \left( \frac{D_{\rm L}}{D_{\rm S} D_{\rm LS}} \right)^{1/2} \;\; ,
\end{equation}
where $D_{\rm LS}$ is the angular diameter distance between the lens L and source S. For example, for $M = 0.01 M_{\odot}$ and a typical source at $z=0.6$ and lens at $z=0.3$, $\eta = 0.50$. As noted in \citet{Pei1993}, ZS17 and \citet{Bosca2022}, the effect of finite sources may be calculated by averaging the point source result over the area of the photosphere. This introduces a maximum magnification $\mu_{\rm max} = \sqrt{1 + 4\eta^2}-1$. Clearly, as $M$ approaches zero, the lensing contribution of compact objects vanishes so that the combined pdf is that of large-scale structure only. 
\par 
To determine whether the finite size of SN Ia sources affects our constraints, we then need to consider the relationship between $\mu_{\rm max}$ (defined by $\eta$ for a reasonable range of redshifts) and the range of $\mu$ of our sample, which we proxy by looking at the distribution of the Hubble diagram residuals (in this subsection, conservatively assuming bright SN Ia are due to magnification, rather than intrinsic brightness).  
\par
\begin{figure}
    \centering
    \includegraphics[width=\columnwidth]{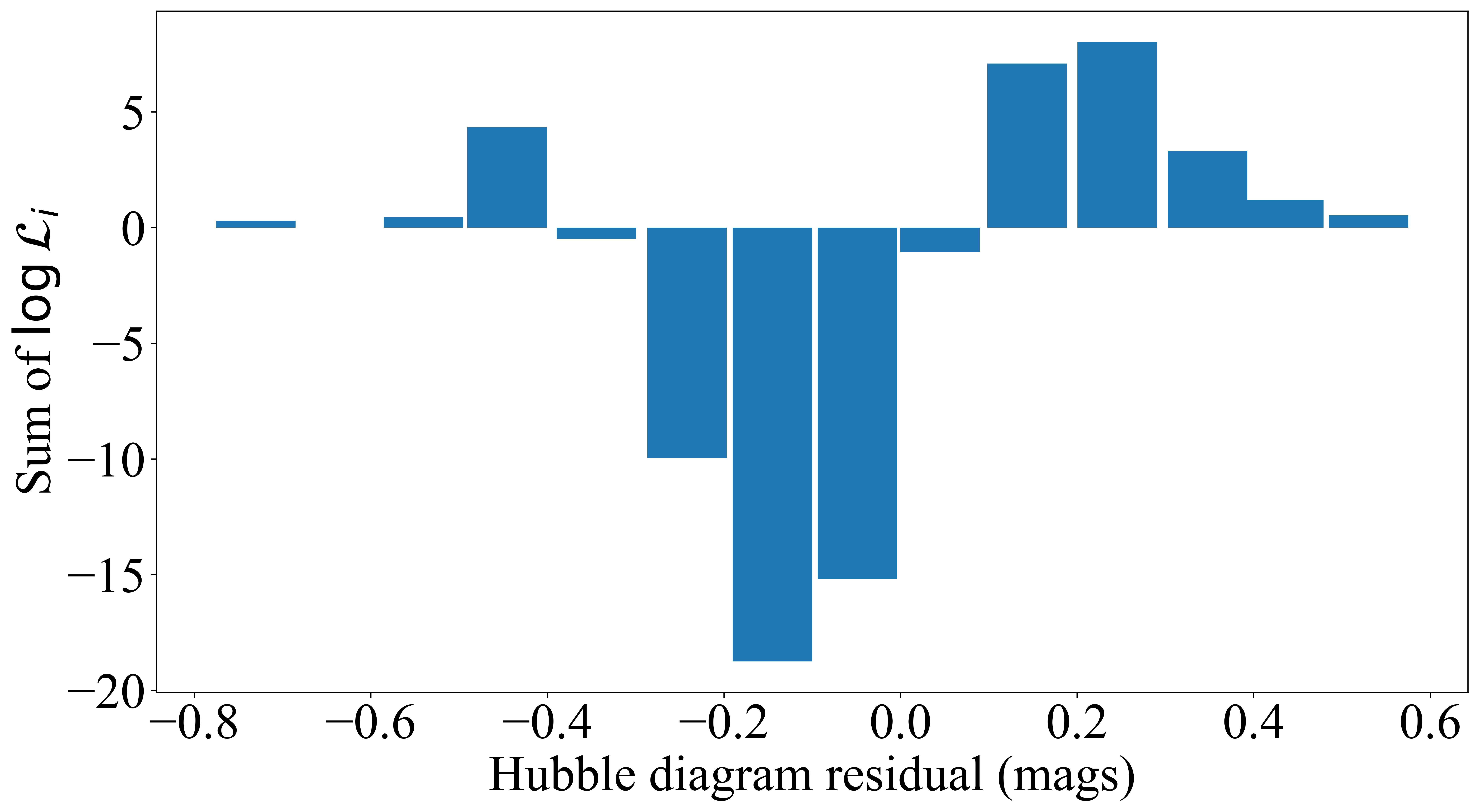}
    \caption{The sum of lensing likelihood contribution per SN Ia for the DES-SN5YR sample, bucketed by Hubble diagram residual. The fraction of compact objects is set at $\alpha=0.3$, and the other parameters used are fiducial values of $\Omega_{\rm m} = 0.315$, $A_s = 2.105 \times 10^{-9}$, $\epsilon = 0$ and $\delta = 1$. This value of $\alpha$ is disfavoured compared to $\alpha =0.01$ at $\Delta \log \mathcal{L} \sim -20$ or $>5 \sigma$. It is clear from the graph this disfavour originates from relative frequency of SNe Ia with residuals of $\Delta m \sim -0.2$ compared to $\Delta m \sim +0.2$.}
    \label{fig:resdist}
\end{figure}

In Figure \ref{fig:resdist}, we show the sum of the lensing likelihood for the DES-SN5YR sample binned by Hubble diagram residual, with $\alpha$ set to $0.3$ and reasonable values for other parameters. For this fraction of compact objects $\delta \log \mathcal{L} \sim -20$ compared to $\alpha=0.01$, meaning that $\alpha=0.3$ is disfavoured compared to the lower value at $>5 \sigma$. We see that the constraints arise from having too many SNe Ia lying just below the Hubble diagram at $\Delta m \sim -0.2$, and not enough just above it, an outcome which is not favoured for lensing due to compact objects. There are no SN Ia lying at $\Delta m < -0.75$ below the Hubble diagram. Using Eqn. \ref{eq:eta} above and $\Delta m \simeq -1.08 \mu$, this corresponds to $\eta = 0.5$. For the range of redshifts spanned by our data and the maximum SN Ia source size, we obtain $0.03 M_{\odot}$ to be an appropriate lower bound to our constraints.
\par
As a further cross-check, Figure 2 of ZS17 shows that below $0.001 M_{\odot}$, the lensing pdf is modified both around the peak and in the tail, whereas above this value it is well-approximated by the point source lensing pdf. Our lower bound of $M > 0.03 M_{\odot}$, is well above $0.001 M_{\odot}$, demonstrating that we can trust our likelihood for this range of compact object masses. 
\par
In summary, the examination of the contribution of individual SN Ia to the lensing likelihood show that the point-source approximation we use is valid for lens masses $M > 0.03 M_{\odot}$. Our constraints do not apply to compact objects lighter than this.

\subsubsection{Non-monochromatic compact object mass}
As the point source pdf does not depend on the compact object mass $M$, a non-monochromatic compact object mass spectrum $p(M) dM$ may be straightforwardly integrated out leaving the results unaffected, provided $p(M)$ has minimal support below  $M> 0.03 M_{\odot}$.
\par
In the case that $p(M)$ does have support for low masses, our constraints would be weaker by a factor $f = P(M< 0.03 M_{\odot})$ where $P$ is the cumulative probability. For example, for a lognormal $p(M)$ with mean $0.03 M_{\odot}$ and width $0.03 M_{\odot}$ then $f=0.69$ and we would have $\alpha < 0.19$ at 95\% confidence.

\subsubsection{PBH clustering}
While it may be assumed that stars and stellar remnants cluster according to the luminosity distribution in galaxies, there is at present a range of proposals in the literature on how PBH may cluster.
\par 
At one extreme, \citet{Farrah2023} has proposed that a  \say{cosmological coupling} of the interior metric of a black hole can explain the origin of dark energy (these are not primordial but remnants from Population III stars). A side effect of the coupling is that the black hole population has an effective equation of state $w=-1$, and the resulting negative effective pressure causes them to disperse out of galactic haloes into the intra-galactic medium. The result is a spatially uniform distribution. Our constraints \textit{do} apply to this scenario, which rules it out as requiring $\alpha \sim \Omega_{\Lambda}/\Omega_{\rm m}$.
\par
On the other hand, the production of PBH as peaks in the density field during inflation has been proposed to generically lead to clustering which survives to the present day \citep{Carr2024}. Cluster sizes can range from $r_c = 1 - 1000$pc and masses from $M_c = 10^3 - 10^6 M_{\odot}$, although a typical scenario seems to be $M_c \sim 10^6 M_{\odot}$ and $r_c  \sim 20$ pc (of similar mass but somewhat smaller than a large globular cluster, and a factor of $\times 1000$ smaller than a smooth dark matter halo of the same mass). Larger clusters are disrupted by Galactic tidal fields, and heavier ones excluded by disc heating \citep[see Figure 5 of][]{Carr2021}. 
\par
It is a well-known theorem that the metric of any spherically-symmetric mass distribution is given by the Schwarzchild metric outside the extent of the mass. The magnification at the Einstein radius as $A = 1+\mu = 1.34$, which as we have noted above is greater than the magnification level from which our results derive. Therefore, if the physical extent of the cluster was contained within its Einstein radius, our result would be robust. For a PBH cluster mass of $10^6 M_{\odot} $, a typical Einstein radius is a physical size of $\sim 10$pc. As this is smaller than the size of the cluster quoted above, our constraint is approximately reduced by the fraction of mass lying outside the Einstein radius. While this clearly depends on the radial profile of the cluster, for a typical NFW profile we may estimate our constraint is reduced to $\alpha < 0.19$ at the 95\% confidence level.
\par
Clearly, clusters that are lower mass or more diffuse can evade our constraints, if they are not disrupted by dynamical evolution considerations. Two scenarios of this would be $M_c = 10^6 M_{\odot}$ and $r_c = 100$pc and $M_c = 10^4 M_{\odot}$ and $r_c = 10$pc. While in principle it would seem a viable scenario that PBH could comprise the entirety of dark matter if they were clustered in this fashion, we note it would be straightforward to re-fashion the lensing pdfs for this scenario and re-run constraints. We do not do this here, and leave it for future consideration. 

\subsubsection{Background cosmology}
\label{sec:backgroundcosmology}
In this paper, we restricted ourselves to Flat $\Lambda$CDM. However, the DES-SN5YR dataset itself indicates a weak preference for \say{thawing} dark energy when combined with other probes (\citet{SNKeyPaper}, see also \citet{DESI2024} who arrived at similar conclusions). However, it has been shown in \citet{Dhawan2023} that the constraint in $\alpha$ varies by $\Delta \alpha/\alpha \sim 0.06$ in extended models (and is tighter for the extended models tested). This is to be expected, as a more flexible Hubble diagram allows the compact object signal to be more clearly de-coupled from the cosmological background. Accordingly, we judge our constraints to be somewhat conservative due to our restriction. 

\section{Summary}
\label{sec:summary}
We have constrained the fraction of compact objects of mass $M > 0.03 M_{\odot}$ as a fraction of the total matter density of the universe to be $\alpha < 0.12$ at the 95\% confidence level. Our results take into account lensing due to the combination of compact objects, large-scale clustering and non-linear scale structure. They apply equally to monochromatic and variable mass spectra of compact objects provided $P(M<0.03 M_{\odot}) \ll 1$. 
\par 
We have argued that our results are biased conservatively (in the sense that the constraints we quote are likely to be higher than the truth) as a result of weak lensing linearisation, the Born approximation, low optical depth, and a Flat $\Lambda$CDM model, in aggregate potentially by a factor of up to 2. We have varied our method for generating the lensing pdf and choices of cuts, finding $\Delta \alpha \sim 0.03$. Our results are robust to differing Hubble expansion rates, and have been fully marginalised over intrinsic non-Gaussianity of residuals and cosmological parameters. Including uncertainty in the validation of the likelihood, we therefore estimate that systematics are limited at $\Delta \alpha \sim 0.04$.
\par
Our constraint on $\alpha$ follows primarily from the lack of clustering of residuals $d_i \sim 0.1$ mag above the Hubble diagram, and also from the lack of SNe Ia with residuals $d_i < -0.5$ mag. Both are signatures of compact objects and distinct from the signature of lensing by large-scale structure which enhances the probability of residuals with $d_i \sim -0.1$ and $d_i \sim +0.05$. 
\par 
Our constraints assume the compact object lenses can be treated as points, and are widely distributed. We are able to rule out the proposal made in \citet{Farrah2023} that \say{cosmologically-coupled} black holes act as the source of dark energy. These black holes originate as stellar remnants, and subsequently disperse widely into the intra-galactic medium (IGM). As such they avoid micro-lensing constraints along LOS confined to our Galactic halo (see next paragraph). However, as SN Ia LOS span large distances over the IGM, our constraints \textit{do} apply to this scenario. For black holes to be the sole source of dark energy would require $\alpha \sim 1$.
\par 
Reviewing other constraints in the literature, the presence of primordial black holes (PBH) in the halo of the Milky Way has been constrained to  $\alpha < 0.012$ in the mass range $1.8 \times 10^{-4} M_{\odot}$ to $6.3 M_{\odot}$ and $\alpha < 0.1$ for masses of $1.3 \times 10^{-5} M_{\odot}$ to $860 M_{\odot}$ at the 95\% confidence level \citep{Mroz2024}, by observing that the frequency of microlensing events between our galaxy and the Large Magellanic Cloud can be almost entirely accounted for by the assumed distribution of halo stars or stellar remnants. The upper limit arises from the duration of the survey, and that microlensing events become rarer as the PBH number density decreases with increasing mass. As our results are based on weak rather than strong lensing, we do not consider there to be an upper limit to our mass range. Nevertheless, as discussed in \citet{Carr2024}, it may be argued that the origin of PBH from primordial peaks in the density field lead to them being strongly clustered into regions of 1-1000pc in size, with cluster masses of $10^3 - 10^6 M_{\odot}$. These scenarios allow the microlensing constraints to be avoided. 
\par
For a reasonable scenario envisoned in \citet{Carr2024}, our result is widened to $\alpha < 0.19$. This is still sufficient to exclude PBH being the sole source of dark matter at high confidence. Although our constraints will become progressively weaker with more diffuse clustering, it would be straightforward to extend our lensing pdf construction to calculate this. We leave this to future work.
\par 
Beyond the halo of our Milky Way, an extragalactic constraint was derived in \citet{Oguri2018} by the observation of the microlensing of a distant star by a foreground massive cluster. It was argued that if the dark matter of the lens was comprised of $\alpha > 0.08 $ compact objects, the smooth caustic would be fragmented to a degree that the observation would not have occurred. The constraint is weakened if a more conservative assumption about stellar velocity dispersions in the foreground cluster is made. In a similar fashion, a constraint of $\alpha < 0.17$ for the mass range $10^4 M_{\odot} < M < 10^6 M_{\odot}$ was obtained by \citet{Dike2023} from the image flux ratios of 11 systems of strongly lensed quasars.
\par
At mass ranges $> 10 M_{\odot}$, tighter constraints than those presented here are argued by a number of lines of reasoning. These include event rates of black hole mergers \citep{AliHamoud2017}, the effect on the CMB power spectrum of accretion onto PBH before the epoch of reionization \citep{Serpico2020}, and number densities of X-ray sources in nearby galaxies such as would be produced by accretion of the inter-stellar medium onto black holes \citep{Inoue2017}. 
\par
Nevertheless, our results are complementary to these in terms of potential systematics and rely only on the action of general relativity to produce lensing magnification. We do not require long timeline observations to collect microlensing events, and our results have no reliance on assumptions about astrophysical processes such as accretion, merger rates and so on. 
\par 
Looking to future avenues of research, it has been proposed that discrepancies between the amplitude of the matter power spectrum derived from the CMB and galaxy-galaxy weak lensing may be due to larger suppression than expected of the power spectrum on scales of the order of 1 Mpc \citep{Amon2022}, possibly due to baryonic effects such as active galactic nucleii and supernovae feedback. We have argued above that SNe Ia provide a unique insight into the power spectrum on these scales, if the fraction of compact objects can be constrained. This will be the subject of a forthcoming paper (in prep.).
\par
The forthcoming Rubin LSST survey is expected to lead to observations of $\mathcal{O}(10^6)$ SNe Ia \citep{Lochner2022}. Since spectroscopic resources will only be available for a fraction of these, the methodology here will need to be adapted to marginalise over uncertain photometric redshifts. Fortunately, due to the weak dependence of the lensing efficiency on the precise redshift of the source and lens, it has been shown that progress may still be made with lensing using photometric samples in S24. A detailed forecast of constraints is beyond the scope of this work. Nevertheless, it may hoped that a definitive \textit{detection} of the presence of compact objects at the level of $\alpha \sim 0.01$ (such as may be expected from stars and stellar remnants alone) may be made.

\section*{Contribution Statement and Acknowledgements}
P.S. devised the project, performed the analysis and drafted the manuscript; M.V. re-analysed the DES-SN5YR data with cuts removed; T.M.D., P.A., L.G., J.L., C.L., O.L., A.M., D.S., M.Su and L.W. advised on the analysis and commented on the manuscript; T.M.D. and O.L. were also internal reviewers and M.Sa was final reader. We particularily thank J.G-B. for extensive comments on earlier versions of the manuscript. The remaining authors have made contributions to this paper that include, but are not limited to, the construction of DECam and other aspects of collecting the data; data processing and calibration; developing broadly used methods, codes, and simulations; running the pipelines and validation tests; and promoting the science analysis.

This paper has gone through internal review by the DES collaboration.

{\footnotesize 
We acknowledge the following former collaborators, who have contributed directly to this work --- Ricard Casas, Pete Challis, Michael Childress, Ricardo Covarrubias, Chris D'Andrea, Alex Filippenko, David Finley, John Fisher, Francisco Förster, Daniel Goldstein, Santiago González-Gaitán, Ravi Gupta, Mario Hamuy, Steve Kuhlmann, James Lasker, Marisa March, John Marriner, Eric Morganson, Jennifer Mosher, Elizabeth Swann, Rollin Thomas, and Rachel Wolf.

T.M.D., A.C., R.C., acknowledge the support of an Australian Research Council Australian Laureate Fellowship (FL180100168) funded by the Australian Government, and A.M. is supported by the ARC Discovery Early Career Researcher Award (DECRA) project number DE230100055.
M.S. and J.L are supported by DOE grant DE-FOA-0002424 and NSF grant AST-2108094.
R.K.\ is supported by DOE grant DE-SC0009924. M.V.\ was partly supported by NASA through the NASA Hubble Fellowship grant HST-HF2-51546.001-A awarded by the Space Telescope Science Institute, which is operated by the Association of Universities for Research in Astronomy, Incorporated, under NASA contract NAS5-26555. 
L.G. acknowledges financial support from the Spanish Ministerio de Ciencia e Innovaci\'on (MCIN), the Agencia Estatal de Investigaci\'on (AEI) 10.13039/501100011033, and the European Social Fund (ESF) ``Investing in your future'' under the 2019 Ram\'on y Cajal program RYC2019-027683-I and the PID2020-115253GA-I00 HOSTFLOWS project, from Centro Superior de Investigaciones Cient\'ificas (CSIC) under the PIE project 20215AT016, and the program Unidad de Excelencia Mar\'ia de Maeztu CEX2020-001058-M, and from the Departament de Recerca i Universitats de la Generalitat de Catalunya through the 2021-SGR-01270 grant. D.S. was supported in part by NASA grant 14-WPS14-0048. The UCSC team is supported in part by NASA grants NNG16PJ34G and NNG17PX03C issued through the Roman Science Investigation Teams Program; NSF grants AST-1518052 and AST-1815935; NASA through grant No. AR-14296 from the Space Telescope Science Institute, which is operated by AURA, Inc., under NASA contract NAS 5-26555; the Gordon and Betty Moore Foundation; the Heising-Simons Foundation.
We acknowledge the University of Chicago’s Research Computing Center for their support of this work.

Funding for the DES Projects has been provided by the U.S. Department of Energy, the U.S. National Science Foundation, the Ministry of Science and Education of Spain, 
the Science and Technology Facilities Council of the United Kingdom, the Higher Education Funding Council for England, the National Center for Supercomputing 
Applications at the University of Illinois at Urbana-Champaign, the Kavli Institute of Cosmological Physics at the University of Chicago, 
the Center for Cosmology and Astro-Particle Physics at the Ohio State University,
the Mitchell Institute for Fundamental Physics and Astronomy at Texas A\&M University, Financiadora de Estudos e Projetos, 
Funda{\c c}{\~a}o Carlos Chagas Filho de Amparo {\`a} Pesquisa do Estado do Rio de Janeiro, Conselho Nacional de Desenvolvimento Cient{\'i}fico e Tecnol{\'o}gico and 
the Minist{\'e}rio da Ci{\^e}ncia, Tecnologia e Inova{\c c}{\~a}o, the Deutsche Forschungsgemeinschaft and the Collaborating Institutions in the Dark Energy Survey.

The Collaborating Institutions are Argonne National Laboratory, the University of California at Santa Cruz, the University of Cambridge, Centro de Investigaciones Energ{\'e}ticas, 
Medioambientales y Tecnol{\'o}gicas-Madrid, the University of Chicago, University College London, the DES-Brazil Consortium, the University of Edinburgh, 
the Eidgen{\"o}ssische Technische Hochschule (ETH) Z{\"u}rich, 
Fermi National Accelerator Laboratory, the University of Illinois at Urbana-Champaign, the Institut de Ci{\`e}ncies de l'Espai (IEEC/CSIC), 
the Institut de F{\'i}sica d'Altes Energies, Lawrence Berkeley National Laboratory, the Ludwig-Maximilians Universit{\"a}t M{\"u}nchen and the associated Excellence Cluster Universe, 
the University of Michigan, NSF's NOIRLab, the University of Nottingham, The Ohio State University, the University of Pennsylvania, the University of Portsmouth, 
SLAC National Accelerator Laboratory, Stanford University, the University of Sussex, Texas A\&M University, and the OzDES Membership Consortium.

Based in part on observations at Cerro Tololo Inter-American Observatory at NSF's NOIRLab (NOIRLab Prop. ID 2012B-0001; PI: J. Frieman), which is managed by the Association of Universities for Research in Astronomy (AURA) under a cooperative agreement with the National Science Foundation.
 Based in part on data acquired at the Anglo-Australian Telescope. We acknowledge the traditional custodians of the land on which the AAT stands, the Gamilaraay people, and pay our respects to elders past and present. Parts of this research were supported by the Australian Research Council, through project numbers CE110001020, FL180100168 and DE230100055. Based in part on observations obtained at the international Gemini Observatory, a program of NSF’s NOIRLab, which is managed by the Association of Universities for Research in Astronomy (AURA) under a cooperative agreement with the National Science Foundation on behalf of the Gemini Observatory partnership: the National Science Foundation (United States), National Research Council (Canada), Agencia Nacional de Investigaci\'{o}n y Desarrollo (Chile), Ministerio de Ciencia, Tecnolog\'{i}a e Innovaci\'{o}n (Argentina), Minist\'{e}rio da Ci\^{e}ncia, Tecnologia, Inova\c{c}\~{o}es e Comunica\c{c}\~{o}es (Brazil), and Korea Astronomy and Space Science Institute (Republic of Korea).  This includes data from programs (GN-2015B-Q-10, GN-2016B-LP-10, GN-2017B-LP-10, GS-2013B-Q-45, GS-2015B-Q-7, GS-2016B-LP-10, GS-2016B-Q-41, and GS-2017B-LP-10; PI Foley).  Some of the data presented herein were obtained at Keck Observatory, which is a private 501(c)3 non-profit organization operated as a scientific partnership among the California Institute of Technology, the University of California, and the National Aeronautics and Space Administration (PIs Foley, Kirshner, and Nugent). The Observatory was made possible by the generous financial support of the W.~M.~Keck Foundation.  This paper includes results based on data gathered with the 6.5 meter Magellan Telescopes located at Las Campanas Observatory, Chile (PI Foley), and the Southern African Large Telescope (SALT) (PIs M.~Smith \& E.~Kasai).
The authors wish to recognize and acknowledge the very significant cultural role and reverence that the summit of Maunakea has always had within the Native Hawaiian community. We are most fortunate to have the opportunity to conduct observations from this mountain.

The DES data management system is supported by the National Science Foundation under Grant Numbers AST-1138766 and AST-1536171.
The DES participants from Spanish institutions are partially supported by MICINN under grants ESP2017-89838, PGC2018-094773, PGC2018-102021, SEV-2016-0588, SEV-2016-0597, and MDM-2015-0509, some of which include ERDF funds from the European Union. IFAE is partially funded by the CERCA program of the Generalitat de Catalunya.
Research leading to these results has received funding from the European Research
Council under the European Union's Seventh Framework Program (FP7/2007-2013) including ERC grant agreements 240672, 291329, and 306478.
We  acknowledge support from the Brazilian Instituto Nacional de Ci\^encia
e Tecnologia (INCT) do e-Universo (CNPq grant 465376/2014-2).

This research used resources of the National Energy Research Scientific Computing Center (NERSC), a U.S. Department of Energy Office of Science User Facility located at Lawrence Berkeley National Laboratory, operated under Contract No. DE-AC02-05CH11231 using NERSC award HEP-ERCAP0023923.
This manuscript has been authored by Fermi Research Alliance, LLC under Contract No. DE-AC02-07CH11359 with the U.S. Department of Energy, Office of Science, Office of High Energy Physics.
} 

\textit{Facilities : }
CTIO:4m, AAT, Gemini:Gillett (GMOS-N), Gemini:South (GMOS-S), Keck:I (LRIS), Keck:II (DEIMOS), Magellan:Baade (IMACS), Magellan:Clay (LDSS3, MagE), SALT

\textit{Software : }
\texttt{numpy} \citep{numpy}, 
\texttt{CAMB} \citep{camb}, 
\texttt{matplotlib} \citep{matplotlib}, 
\texttt{scipy} \citep{scipy}, 
\texttt{SNANA} \citep{Snana}, 
\texttt{Pippin} \citep{Pippin}, 
\texttt{Polychord} \citep{Polychord}, 
\texttt{Anesthetic} \citep{Anesthetic},
\texttt{PlikLite} Python implementation \citep{prince19},
\texttt{HMCODE2020} \citep{Mead2020},
\texttt{TurboGL} \citep{Kainulainen2009}

\section*{Data Availability}
The data and Python code used to generate the results and plots in this paper are available on reasonable request from the authors. The lensing pdfs used will be made available upon publication at \url{https://github.com/paulshah/SNLensing}.
 



\bibliographystyle{mnras}
\bibliography{DEScitations} 



\appendix
\section{Results validation}
\label{sec:validation}
We validate our results by generating mock datasets of similar size to DES-SN5YR using the software packages \texttt{SNANA}\footnote{\url{https://github.com/RickKessler/SNANA}} \citep{Snana} and \texttt{PIPPIN}\footnote{\url{https://github.com/dessn/Pippin}} \citep{Pippin}. The mock datasets replicate the observing conditions and detection efficiency of the DES Survey, which have been used to analyse the actual observations. The simulations allow the user to specify the lensing pdf to use, and we select a number of values of $\alpha$, generate 10 mock datasets for each, and analyse them using our likelihood. 
\par
\begin{figure}
    \centering
    \includegraphics[width=\columnwidth]{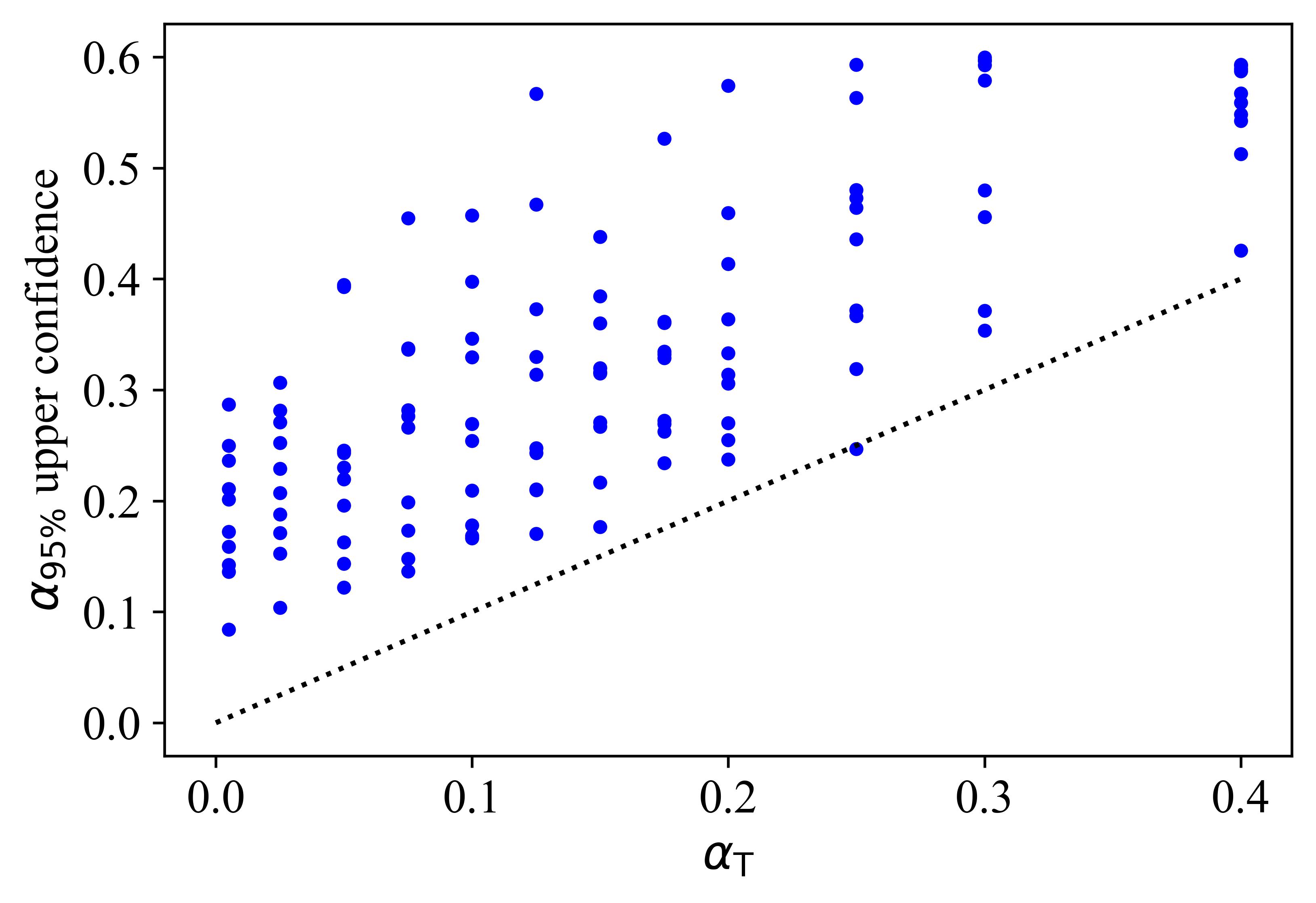}
    \caption{Upper confidence interval statistics for each of ten mock data sets for values of $\alpha$ from 0.005 to 0.40. The truth is plotted as a diagonal dashed black line. The mock datasets are generated by SNANA simulations with observing conditions matching the real survey, but with weak lensing magnifications sampled from our model pdfs. For each simulation, we plot the 95\% upper confidence limit statistic using CMB priors. The trend on this statistic is somewhat flatter than the truth, as is evident due to the hard prior boundaries. These data points are then used, via Bayes' theorem, to validate that our likelihood correctly retrieves the limit statistic.}
    \label{fig:sims}
\end{figure}

 
Our objective is to validate our 95\% confidence upper limit for $\alpha$.  We note that any statistic will be necessarily bounded by our prior interval $\alpha \in (0,0.7)$, and therefore is likely to be biased for a truth value $\alpha_{\rm T}$ close to either side of this interval (for example, if we create multiple simulations with $\alpha_{\rm T} = 0 $, all statistics will be higher than this by construction). Accordingly, we examine the distribution $p(\alpha_{95\%} | \alpha_{\rm T})$ where $\alpha_{95\%}$ is the 95\% confidence limit from the simulated data with input lensing pdf $\alpha = \alpha_{\rm T}$ and fiducial values for $A_s$ and $\Omega_{\rm m}$. The results with CMB priors are shown in Figure \ref{fig:sims} \footnote{These results were derived using lower resolution Polychord settings due to compute time needed. We have checked in a few cases they adequately replicate the results of our standard settings.}.
\par
We see that for only one out of 120 of our simulations does the simulated value fall lower than the 95\% confidence upper limit. The trend on the statistic is somewhat flatter than the truth, due to the hard prior boundaries. We may use Bayes' theorem to construct $p(\alpha_{\rm T} | \alpha_{95\%})$ by fitting a kernel density estimate to $p(\alpha_{\rm T} , \alpha_{95\%})$, assuming a flat prior on $p(\alpha_{\rm T})$, and integrating to find $p(\alpha_{\rm T} < \alpha_{95\%} | \alpha_{95\%})$. For the value $\alpha_{95\%} = 0.14$ we recover $95\%$, indicating our likelihood is accurate for our quoted result to within $\Delta \alpha \sim 0.02$.


\noindent \\
$^{1}$ Department of Physics \& Astronomy, University College London, Gower Street, London, WC1E 6BT, UK\\
$^{2}$ School of Mathematics and Physics, University of Queensland,  Brisbane, QLD 4072, Australia\\
$^{3}$ Department of Physics, University of Oxford, Denys Wilkinson Building, Keble Road, Oxford OX1 3RH, UK\\
$^{4}$ School of Physics and Astronomy, University of Southampton,  Southampton, SO17 1BJ, UK\\
$^{5}$ The Research School of Astronomy and Astrophysics, Australian National University, ACT 2601, Australia\\
$^{7}$ Center for Astrophysics $\vert$ Harvard \& Smithsonian, 60 Garden Street, Cambridge, MA 02138, USA\\
$^{8}$ Institut d'Estudis Espacials de Catalunya (IEEC), 08034 Barcelona, Spain\\
$^{9}$ Institute of Space Sciences (ICE, CSIC),  Campus UAB, Carrer de Can Magrans, s/n,  08193 Barcelona, Spain\\
$^{10}$ Instituto de Fisica Teorica UAM/CSIC, Universidad Autonoma de Madrid, 28049 Madrid, Spain\\
$^{11}$ SLAC National Accelerator Laboratory, Menlo Park, CA 94025, USA\\
$^{12}$ Department of Physics and Astronomy, University of Pennsylvania, Philadelphia, PA 19104, USA\\
$^{13}$ Centre for Gravitational Astrophysics, College of Science, The Australian National University, ACT 2601, Australia\\
$^{14}$ Centre for Astrophysics \& Supercomputing, Swinburne University of Technology, Victoria 3122, Australia\\
$^{15}$ Department of Physics, Duke University Durham, NC 27708, USA\\
$^{16}$ Universit\'e Grenoble Alpes, CNRS, LPSC-IN2P3, 38000 Grenoble, France\\
$^{17}$ Fermi National Accelerator Laboratory, P. O. Box 500, Batavia, IL 60510, USA\\
$^{18}$ Laborat\'orio Interinstitucional de e-Astronomia - LIneA, Rua Gal. Jos\'e Cristino 77, Rio de Janeiro, RJ - 20921-400, Brazil\\
$^{19}$ University Observatory, Faculty of Physics, Ludwig-Maximilians-Universit\"at, Scheinerstr. 1, 81679 Munich, Germany\\
$^{20}$ Kavli Institute for Particle Astrophysics \& Cosmology, P. O. Box 2450, Stanford University, Stanford, CA 94305, USA\\
$^{21}$ Instituto de Astrofisica de Canarias, E-38205 La Laguna, Tenerife, Spain\\
$^{22}$ Hamburger Sternwarte, Universit\"{a}t Hamburg, Gojenbergsweg 112, 21029 Hamburg, Germany\\
$^{23}$ Department of Physics, IIT Hyderabad, Kandi, Telangana 502285, India\\
$^{24}$ Department of Physics, Carnegie Mellon University, Pittsburgh, Pennsylvania 15312, USA\\
$^{25}$ NSF AI Planning Institute for Physics of the Future, Carnegie Mellon University, Pittsburgh, PA 15213, USA\\
$^{26}$ Institute of Theoretical Astrophysics, University of Oslo. P.O. Box 1029 Blindern, NO-0315 Oslo, Norway\\
$^{27}$ Kavli Institute for Cosmological Physics, University of Chicago, Chicago, IL 60637, USA\\
$^{28}$ Center for Astrophysical Surveys, National Center for Supercomputing Applications, 1205 West Clark St., Urbana, IL 61801, USA\\
$^{29}$ Department of Astronomy, University of Illinois at Urbana-Champaign, 1002 W. Green Street, Urbana, IL 61801, USA\\
$^{30}$ Santa Cruz Institute for Particle Physics, Santa Cruz, CA 95064, USA\\
$^{31}$ Center for Cosmology and Astro-Particle Physics, The Ohio State University, Columbus, OH 43210, USA\\
$^{32}$ Department of Physics, The Ohio State University, Columbus, OH 43210, USA\\
$^{33}$ Australian Astronomical Optics, Macquarie University, North Ryde, NSW 2113, Australia\\
$^{34}$ Lowell Observatory, 1400 Mars Hill Rd, Flagstaff, AZ 86001, USA\\
$^{35}$ Jet Propulsion Laboratory, California Institute of Technology, 4800 Oak Grove Dr., Pasadena, CA 91109, USA\\
$^{36}$ George P. and Cynthia Woods Mitchell Institute for Fundamental Physics and Astronomy, and Department of Physics and Astronomy, Texas A\&M University, College Station, TX 77843,  USA\\
$^{37}$ LPSC Grenoble - 53, Avenue des Martyrs 38026 Grenoble, France\\
$^{38}$ Instituci\'o Catalana de Recerca i Estudis Avan\c{c}ats, E-08010 Barcelona, Spain\\
$^{39}$ Institut de F\'{\i}sica d'Altes Energies (IFAE), The Barcelona Institute of Science and Technology, Campus UAB, 08193 Bellaterra (Barcelona) Spain\\
$^{40}$ Department of Astrophysical Sciences, Princeton University, Peyton Hall, Princeton, NJ 08544, USA\\
$^{41}$ Observat\'orio Nacional, Rua Gal. Jos\'e Cristino 77, Rio de Janeiro, RJ - 20921-400, Brazil\\
$^{42}$ Department of Physics, Northeastern University, Boston, MA 02115, USA\\
$^{43}$ Centro de Investigaciones Energ\'eticas, Medioambientales y Tecnol\'ogicas (CIEMAT), Madrid, Spain\\
$^{44}$ Physics Department, Lancaster University, Lancaster, LA1 4YB, UK\\
$^{45}$ Computer Science and Mathematics Division, Oak Ridge National Laboratory, Oak Ridge, TN 37831\\
$^{46}$ Department of Physics, University of Michigan, Ann Arbor, MI 48109, USA\\
$^{47}$ Department of Astronomy, University of California, Berkeley,  501 Campbell Hall, Berkeley, CA 94720, USA\\
$^{48}$ Lawrence Berkeley National Laboratory, 1 Cyclotron Road, Berkeley, CA 94720, USA\\

\bsp	
\label{lastpage}
\end{document}